\documentclass[pra,twocolumn,preprintnumbers ,amsmath, amssymb, superscriptaddress, aps]{revtex4-2}

\usepackage{color}
\usepackage{amsmath}
\usepackage{pifont}
\usepackage{amssymb}  
\usepackage{bbold}
\usepackage{float}
\usepackage{tikz}
\usepackage{makecell}
\usepackage{subfigure}
\usepackage{pifont}   
\usepackage{graphicx} 
\usepackage{dcolumn}  
\usepackage{bm}       
\usepackage{multirow} 
\usepackage{hyperref}
\usepackage{placeins}
\newcommand{\vect}[1]{\mathbf{#1}}

\begin{document}

\renewcommand\floatpagefraction{0.8} 
\renewcommand\topfraction{0.8}       

\author{Michael Vogl}
\thanks{These two authors contributed equally.}
\affiliation{Department of Physics, King Fahd University of Petroleum and Minerals, 31261 Dhahran, Saudi Arabia}
\author{Martin Rodriguez-Vega}
\thanks{These two authors contributed equally.}
\affiliation{Department of Physics, The University of Texas at Austin, Austin, TX 78712, USA}
\affiliation{Department of Physics, Northeastern University, Boston, MA 02115, USA}
\author{Benedetta Flebus}
\affiliation{Department of Physics, The University of Texas at Austin, Austin, TX 78712, USA}
\affiliation{Department of Physics, Boston College, 140 Commonwealth Avenue, Chestnut Hill, MA 02467, USA}
\author{Allan H. MacDonald}
\affiliation{Department of Physics, The University of Texas at Austin, Austin, TX 78712, USA}
\author{Gregory A. Fiete}
\affiliation{Department of Physics, Northeastern University, Boston, MA 02115, USA}
\affiliation{Department of Physics, Massachusetts Institute of Technology, Cambridge, MA 02139, USA}
\title{Floquet-engineering topological transitions in a twisted transition metal dichalcogenide homobilayer}
\date{\today}

\begin{abstract}
%
Motivated by the recent experimental realization of twisted transition metal dichalcogenide bilayers,  we study a simplified model driven by different forms of monochromatic light. As a concrete and representative example we use parameters that correspond to a twisted MoTe$_2$ homobilayer. First, we consider irradiation with circularly polarized light in free space and demonstrate that the corresponding Floquet Hamiltonian takes the same form as the static Hamiltonian, only with a constant overall shift in quasi-energy. This is in stark contrast to twisted bilayer graphene, where new terms are typically generated under an analagous drive. Longitudinal light, on the other hand, which can be generated from the transverse magnetic mode in a waveguide, has a much more dramatic effect--it renormalizes the tunneling strength between the layers, which effectively permits the tuning of the twist angle {\em in-situ}. We find that, by varying the frequency and amplitude of the drive, one can induce a topological transition that cannot be obtained with the traditional form of the Floquet drive in free space. Furthermore, we find that strong drives can have a profound effect on the layer pseudo-spin texture of the twisted system, which coincides with multiple simultaneous band gap closings in the infinite-frequency limit. Surprisingly, these bandgap closings are not associated with topological transitions. For high but finite drive frequencies near $0.7$eV, the infinite-frequency band crossings become band gap minima of the order of $10^{-6}$ eV or smaller. 
\end{abstract}
\maketitle

\section{Introduction}

 After the discovery of superconductivity in twisted bilayer graphene \cite{cao2018unconventional} there has been a tremendous interest in moir\'e materials \cite{Wu_2018,Bistritzer12233,Cao2018sc,Codecidoeaaw9770,Yankowitz1059,chichinadze2019nematic,Chou_2019,Guinea13174,PhysRevLett.122.257002,PhysRevB.99.134515,caldern2019correlated,saito2019decoupling,stepanov2019interplay,Kang_2019,Volovik_2018,Po_2018,Ochi_2018,Gonz_lez_2019,Sherkunov_2018,Laksono_2018,Venderbos_2018,PhysRevB.81.165105,Seo_2019,PhysRevB.93.035452,Cao2018,Kim3364,Sharpe605,PhysRevB.98.085144,Chittari_2018,Yankowitz2018,shang2019artificial,fleischmann2019moir,May_Mann_2020,Wu_2019,zhai2020theory,zhang2020tuning,zhan2020multiultraflatbands,venkateswarlu2020electronic,ruiztijerina2020theory,chen2020configure}. Particularly notable are those derived from graphene such as twisted bilayers \cite{Wu_2018,Bistritzer12233,Cao2018sc,Codecidoeaaw9770,Yankowitz1059,chichinadze2019nematic,Chou_2019,Guinea13174,PhysRevLett.122.257002,PhysRevB.99.134515,caldern2019correlated,saito2019decoupling,stepanov2019interplay,Kang_2019,Volovik_2018,Po_2018,Ochi_2018,Gonz_lez_2019,Sherkunov_2018,Laksono_2018,Venderbos_2018,PhysRevB.81.165105,Seo_2019,PhysRevB.93.035452,Cao2018,Kim3364,Sharpe605,PhysRevB.98.085144,Chittari_2018,Yankowitz2018,shang2019artificial,fleischmann2019moir}, and stretched and strained graphene bilayers \cite{San_Jose_2012,PhysRevB.96.035442}. Much of this interest is derived from the possibility to engineer moir\'e patterns that lead to a rich twist angle-dependent band structure. Furthermore one finds flat bands at specific magic angles \cite{Bistritzer12233}.  Flat bands imply small to vanishing kinetic energy, which renders the electron-electron interactions a dominant energy scale. Therefore, flat band systems can host strongly correlated phases of matter and bilayers of various twist angles offer an opportunity to tune the band flatness.
 
An interesting class of materials closely related to graphene is transition metal dichalcogenides (TMDs). TMDs are a family of materials with chemical formula $\text{MX}_2$, where $\text{M}$ is a transition metal (e.g. $\text{W}$ or $\text{Mo}$) and $\text{X}$ any of the three chalcogens $\text{S}$, $\text{Se}$ or $\text{Te}$ \cite{manzeli20172d,wang2012electronics}. They have recently attracted attention because of their interesting electronic and optical properties \cite{wang2012electronics}. Some of this interest is owed to their two-dimensional nature and some to their direct band gap with associated frequency in the optical range. TMDs lend themselves to the design of electronic components such as transistors \cite{xu2014spin,wang2012electronics}. Furthermore they display interesting optical effects such as an extraordinary large value of the refractive index in the visible frequency range \cite{liu2014optical}. Due to their dimensionality and lattice structure, the study of moir\'e superlattices has been extended to TMDs. Theoretical work \cite{Wu_2019,zhai2020theory,zhang2020tuning,zhan2020multiultraflatbands,venkateswarlu2020electronic} has proposed interesting effects in different twisted TMDs bilayers (tTMDs) including  flatbands appearing for a range of angles \cite{zhan2020multiultraflatbands} and angle dependent topological transitions \cite{Wu_2019}. Experimentally, evidence for moir\'e excitons has been reported in tTMD heterobilayers \cite{Tran2019,Seyler2019,Alexeev2019}.

A second line of research with recent rapid development is periodically driven quantum systems or Floquet-systems, e.g., a system irradiated by monochromatic laser light. Such studies have been motivated by the prediction of phases of matter and transitions that cannot be achieved at equilibrium \cite{PhysRevLett.116.250401,PhysRevX.6.041001,Yao2017a,PhysRevLett.119.123601,Lerose_2019}, as well as the possibility to engineer rich topological structures \cite{wang2013,PhysRevLett.106.220402,PhysRevB.96.125144,PhysRevLett.114.125301,PhysRevLett.116.176401,PhysRevB.95.035136,PhysRevB.97.035422,PhysRevX.6.021013,DAlessio2015,PhysRevX.7.041008,PhysRevB.79.081406,Lindner2011,PhysRevB.84.235108,PhysRevLett.107.216601,PhysRevB.97.035416,PhysRevX.3.031005,PhysRevB.90.115423}. This line of research is also motivated by the possibility of studying these periodically driven systems by effective time-independent Floquet Hamiltonians, which greatly simplify their treatment. Several theoretical approaches have been developed and include perturbative methods and their applications \cite{Abanin_2017,Eckardt_2015,Bukov_2015,Mikami_2016,Blanes2009,Magnus1954,PhysRevA.68.013820,PhysRevX.4.031027,PhysRevB.97.155434,PhysRevLett.115.075301,PhysRevB.94.235419,PhysRevLett.116.125301,PhysRevB.25.6622,Martiskainen2015,Rodriguez_Vega_2018,rodriguezvega2020moirefloquet} and non-pertubative methods \cite{Vajna_2018,Bukov_2015,Vogl_2020,PhysRevA.7.2203,Vogl_2019,PhysRevLett.111.175301,rodriguezvega2020moirefloquet,Vogl_2019Flow}. Inevitably, Floquet studies have appeared for graphene \cite{PhysRevB.79.081406,PhysRevB.95.125401,PhysRevB.93.115420,kristinsson2016control,PhysRevB.81.165433,PhysRevB.89.121401,PhysRevB.90.115423,PhysRevA.91.043625,mciver2020light,Vogl_2020,Lindner2011}. In graphene one may apply circularly polarized light to open a gap and break time-reversal symmetry. Similarly, there have been studies on TMDs under the influence of light \cite{huaman2019floquet,Sie_2014,sengupta2016photo}; circularly polarized light can lift the valley degeneracy \cite{Sie_2014}.

 More recently there has been interest in the effects of light on moir\'e materials. For instance, there have been multiple studies of twisted bilayer graphene subjected to circularly polarized light \cite{Vogl_2020floq,katz2019optically,Topp_2019,li2019floquetengineered,PhysRevResearch.2.032015} and transverse magnetic mode (longitudinal component) light, like the transverse magnetic mode from a waveguide \cite{Vogl_2020interlayer}. Twisted double bilayer graphene \cite{rodriguezvega2020floquet} has been studied under the influence of both forms of light. A recent review of these topics can be found at \cite{rodriguezvega2020moirefloquet}. However, for tTMDs these types of monochromatic drives have not been investigated yet.  This is the topic of this work. 
 
The remainder of the paper is organized as follows. In Sec. \ref{sec:undriven} we give a brief summary of relevant results regarding the undriven tTMD and its effective Hamiltonian. In Sec. \ref{sec:regularize} we discuss the limitation the undriven model faces when subjected to periodic drive and how the model has to be regularized to accurately capture the influence of a periodic drive. In Sec. \ref{sec:coupl} we discuss how light of different polarizations (including longitudinal) couples in the effective model and provide a brief discussion of the numerical implementation of the time-dependent Hamiltonians. The effects of light on band topology, pseudo-spin texture, and bandstructure we observe in driven systems and their effective time-independent description are discussed in Sec. \ref{sec:phys_imp_liights}. Lastly we summarize our results in Sec. \ref{sec:conculsions}.

\section{The undriven model}
\label{sec:undriven}

\begin{figure}[t]
	\centering
	\includegraphics[width=1\linewidth]{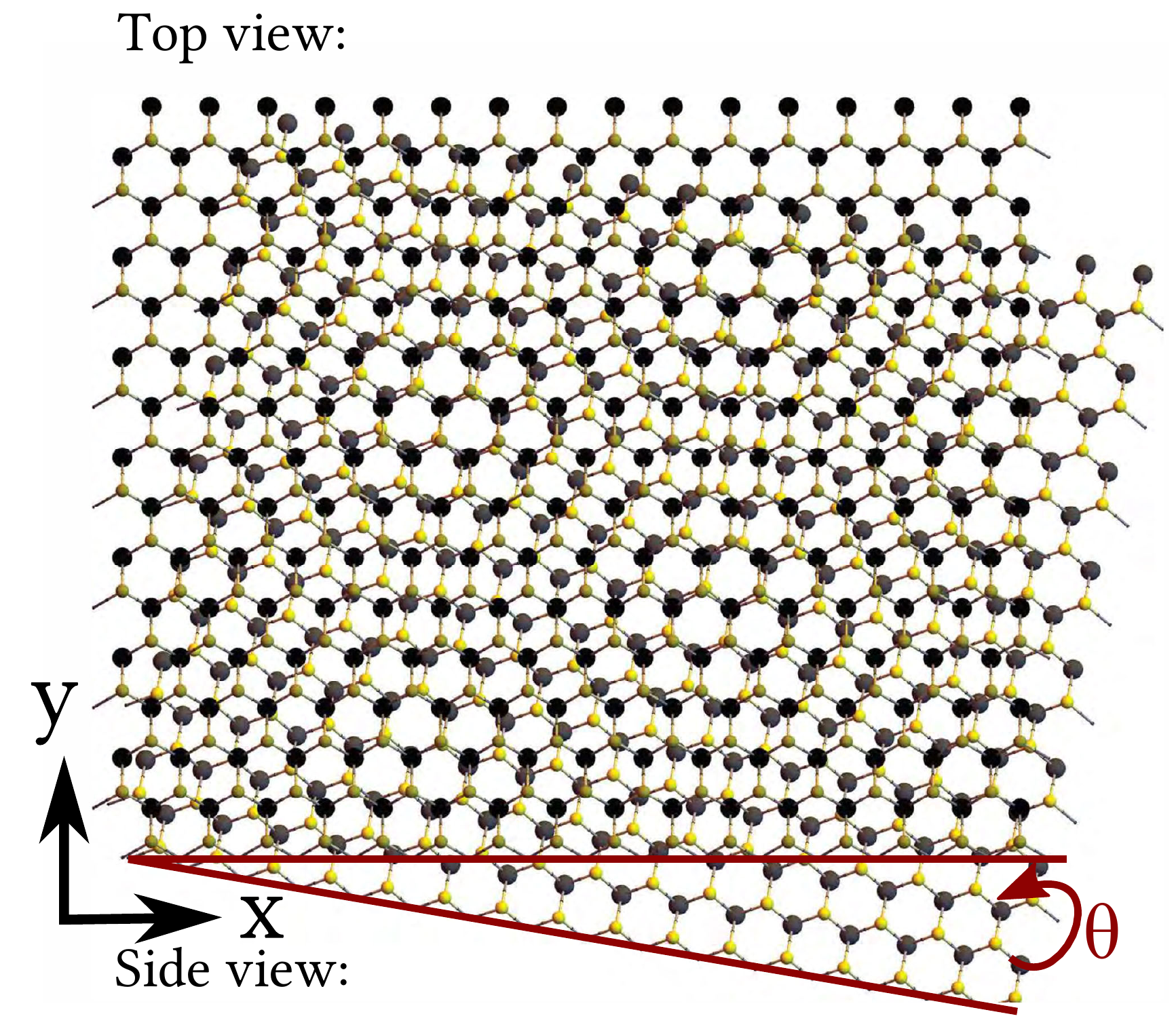}
	\includegraphics[width=1\linewidth]{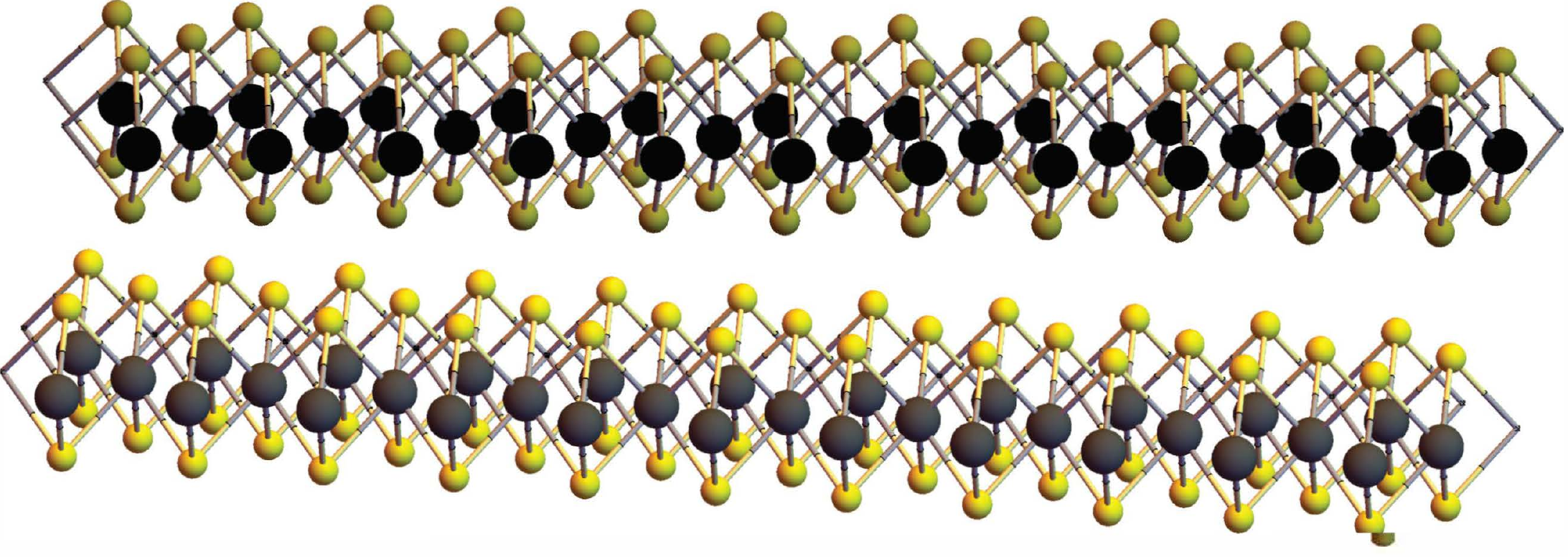}
	\caption{(Color online) Top and side view of a twisted TMD double layer. The black atoms correspond to the top layer $Mo$ atoms, the gray atoms to the bottom layer. The dark yellow atoms correspond to top layer $Te$ atoms and the yellow ones to the bottom layer.}
	\label{fig:topviewtwistmose2bilayer}
\end{figure}

The starting point of our discussion for a twisted TMD homobilayer, as shown in Fig. \ref{fig:topviewtwistmose2bilayer} for $\text{MoTe}_2$, is the low-energy effective Hamiltonian  introduced in Ref.~\cite{Wu_2019}
\begin{equation}
\hspace*{-0.245cm}
\mathcal{H}_{\uparrow}( \vect r)[f]=\begin{pmatrix}
f(\vect k-\kappa_+) +\Delta_{1}(\vect r) & \Delta_{T}(\vect r) \\
\Delta_{T}^{\dagger}(\vect r) & f(\vect k-\kappa_-) +\Delta_{-1}(\vect r)
\end{pmatrix},
\label{Hd0}
\end{equation}
where $f(\vect k)=-\vect k^2/(2 m^*)$ is the approximate low-energy valence band dispersion of a single layer TMD near the $\vect K$ point. We will later introduce an alternate approximation for this term since a bounded Hamiltonian is more appropriate in a driven setting because it improves convergence properties. Alternatively, one could introduce a cut-off energy.

The Hamiltonian Eq.\eqref{Hd0} is defined in the basis $\Psi=(\Psi_b,\Psi_t)$, where $\Psi_b$ ($\Psi_t$) corresponds to the bottom (top) layer creation operator with spin up. The two layers are mutually rotated with respect to each other by an angle $\theta$, which will be assumed to be small, i.e., $\theta\lesssim 10^\circ$. A bilayer with larger twist angles becomes quasi-periodic and therefore will not be considered here. To have momenta in both layers measured with respect to a common coordinate system one needs to introduce the mutual shifts $\kappa_{\pm}=2 \pi  \theta  (-1/\sqrt{3},\pm1/3)/a_0$, where $a_0$ is the intra layer lattice constant. 
 
\begin{figure}
	\centering
	\includegraphics[width=1\linewidth]{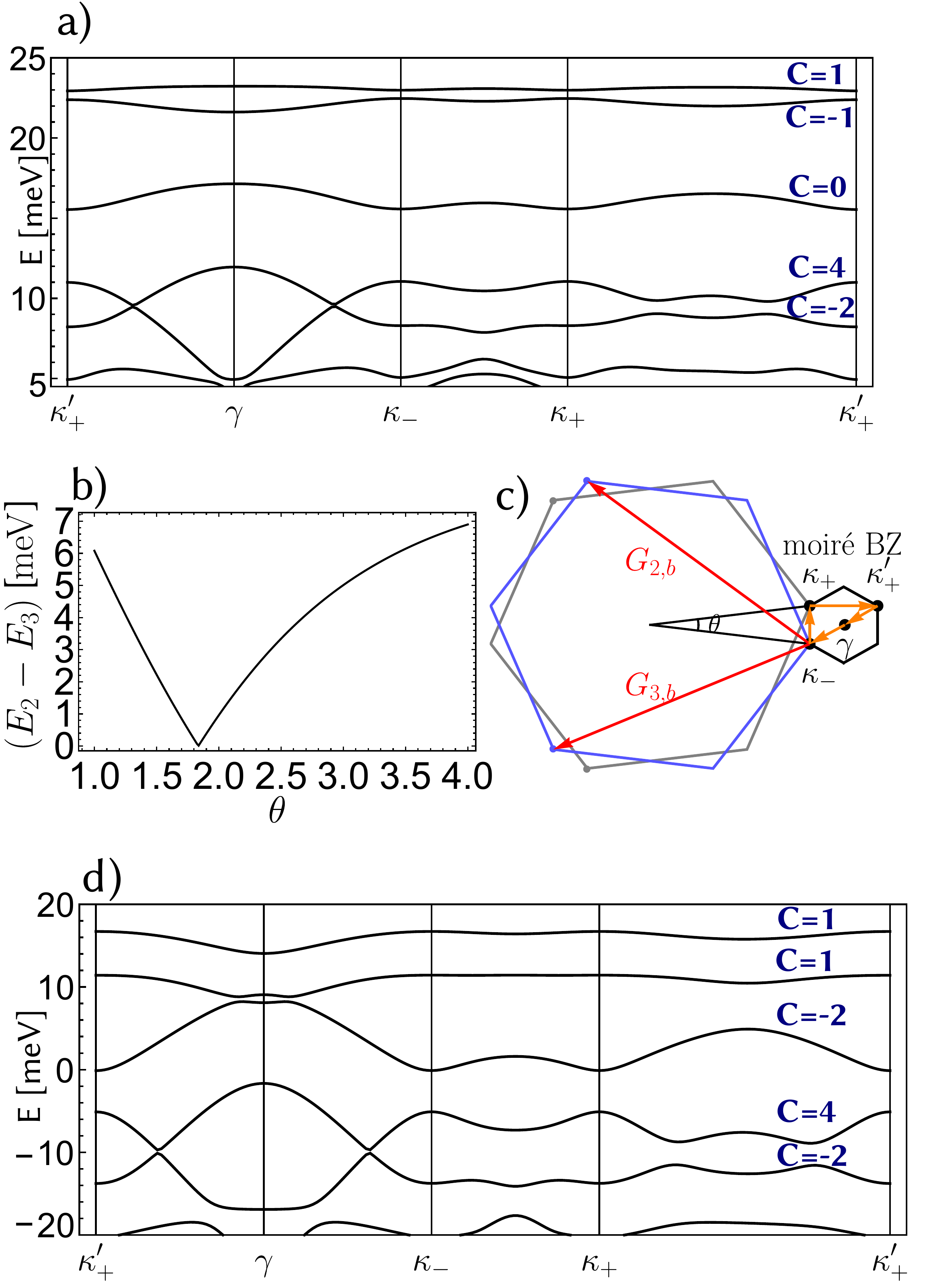}
\caption{(Color online) The figure shows how a variation of twist angle can induce a band gap closing and reopening and how Chern numbers of the bands change as a result. In subfigure a), the plot shows the energy spectrum for a twist angle $1.2^\circ$ along a high symmetry path in the Brillouin zone; in subfigure b) we show the band gap between the second band $E_2$ and third band $E_3$ and find that it closes at an angle of approximately $1.8^\circ$; and in subfigure d) we show the energy spectrum along a high symmetry path for a twist angle of  $2^\circ$. Panel c) shows the moir\'e BZ. It should be noted that the gaps between bands four and five are small of the order of $\sim0.1$ meV and therefore are difficult to see in the figures.}
\label{fig:undriven1p2deg}
\end{figure}

Geometrically, the rotational mismatch between the layers leads to a moir\'e pattern that is captured by the following effective interlayer tunneling term
\begin{equation}
\Delta_{T}(\vect r) = w (1+ e^{-i \theta\vect G_2 \cdot (\hat z\times \vect r)}+ e^{-i \theta\vect G_3 \cdot (\hat z\times \vect r)}),
\label{Tunneling}
\end{equation}
where $\vect G_n=4\pi/(\sqrt{3}a_0) R_z((n-1)\pi/3)\hat y$ and $R_z$ is a rotation matrix around the $z$ axis. Here, $w$ determines the strength of the interlayer coupling. Furthermore, we have an effective position-dependent layer bias that is given as
\begin{equation}
	\Delta_l=2V\sum_{j=1}^3\cos(\theta\vect G_{2j+1}(\hat z\times \vect r)+l\psi),
	\label{delta_l}
\end{equation}
where $V$ sets the strength of the position-dependent in-plane bias and the index $l=\pm 1$. These position dependent terms in the Hamiltonian correspond to the moir\'e potential, and endow the twisted TMD with a larger unit cell than a single layer TMD. The characteristic length scale is $a_M=a_0/\theta$, and has an associated smaller moir\'e Brillouin zone (mBZ) as seen in Fig.\ref{fig:undriven1p2deg}. Through out this work and for concreteness, we fix the physical parameters to ones that correspond to MoTe$_2$. That is we consider the model as in Ref. [\onlinecite{Wu_2019}], i.e., 
$
	(V,w,\psi,m^*,a_0)=(8 \text{meV},-8.5 \text{meV},-89.6^\circ,0.62m_e,3.47 \text{\AA}),
$
where $m^*$ is an effective mass, $\psi$ a phase term and $a_0$ is the intra-layer distance between sites. 

We will now review some of the interesting equilibrium properties of Hamiltonian that were discussed in Ref.   \cite{Wu_2019}. It is useful to define a layer-space pseudo-spin magnetic field as
\begin{equation}
	\vect \Delta(\vect r)=\left(\mathrm{Re}(\Delta_T^\dag(\vect r)),\mathrm{Im}(\Delta_T^\dag(\vect r)),\frac{\Delta_1(\vect r)-\Delta_{-1}(\vect r)}{2}\right)^T.
	\label{layer_pseudo_Spin}
\end{equation}

\begin{figure}[]
	\centering
	\includegraphics[width=1\linewidth]{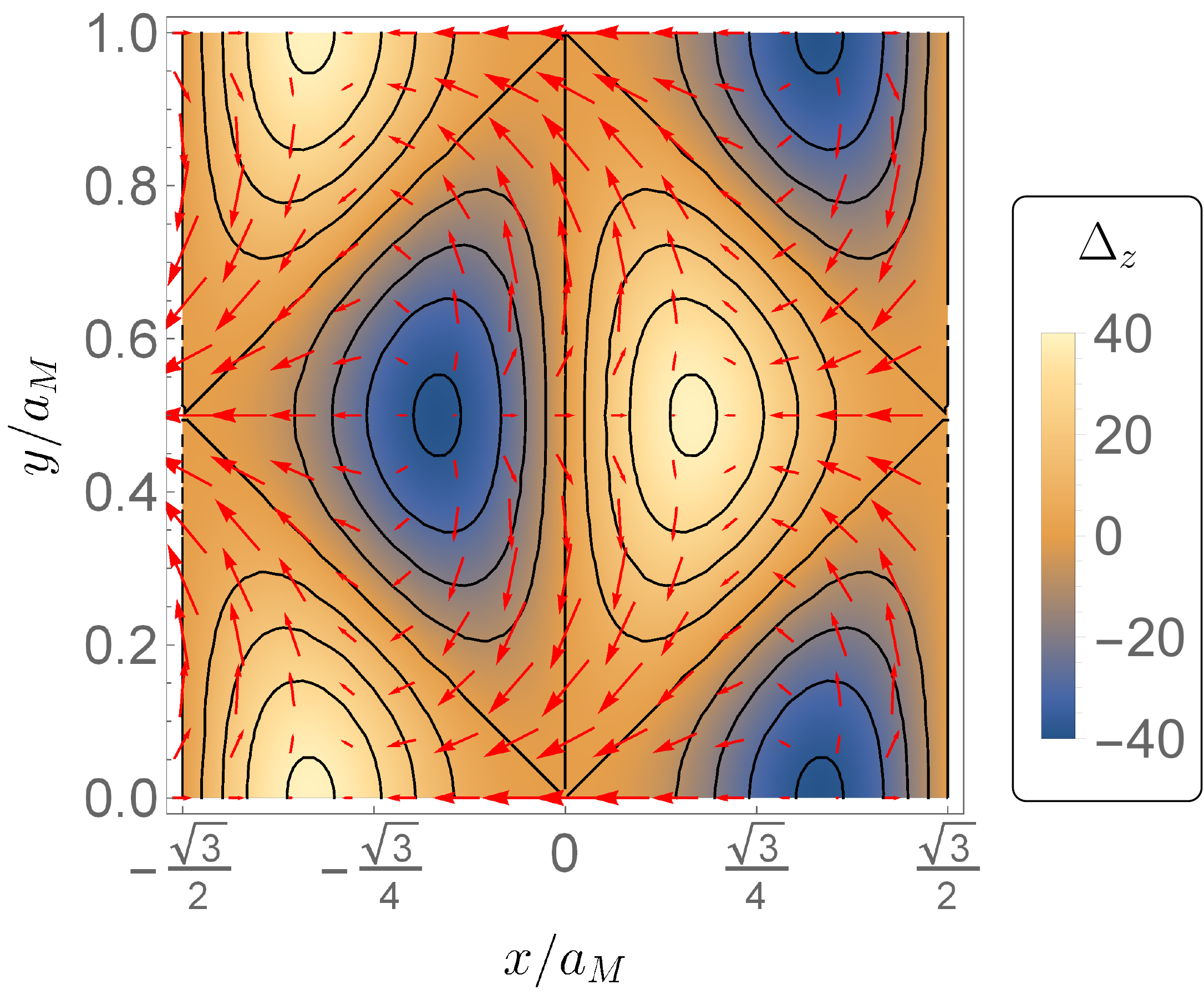}
	\caption{(Color online) Plot of $\vect \Delta(\vect r)=(\Delta_{x},\Delta_{y},\Delta_z)$, where $\Delta_{x,y}$ correspond to the red arrows and the density plot corresponds to $\Delta_z$.}
	\label{fig:skyrmions}
\end{figure}

As shown in Fig. \ref{fig:skyrmions}, for this system, $\vect \Delta(\vect r)$ exhibits a layer space skyrmion pseudo-spin texture, which winds once around the moir\'e unit cell \cite{Wu_2019, nagaosa2013topological}.  

In Fig. \ref{fig:undriven1p2deg} (a) we plot the band structure for a twist angle of $1.2^\circ$. The top three bands exhibit a narrow bandwidth. Furthermore, the different bands have non-trivial Chern numbers (as indicated in the right of the figure)
	\begin{equation}
	C_n = \frac{1}{2 \pi i} \int_{\text{mBZ}} \left( \nabla \times \langle u_{n}( \mathbf{k})\left|\partial_{\vect k}\right| u_{n}(\mathbf{k})\rangle \right)_zd \vect k,
	\label{eq:chern}
	\end{equation}
where $|u_{n}\rangle$ is the $n$-th eigenvector of the Bloch Hamiltonian. As the twist angle is increased, the gap between bands $n=2$ and $n=3$ decreases (counting from the ``top"), until the gap closes at $\theta \approx 1.8^\circ$, as shown in Fig. \ref{fig:undriven1p2deg} (c). This gap closing is accompanied by a change in the band Chern numbers from $(-1,0)$ to $(1,-2)$ (Fig. \ref{fig:undriven1p2deg} (d)).

It is interesting that such a topological transition may be achieved merely by changing the angle.  However, in order to study the physics across the transition, one would have to prepare a new sample for each angle and match the angle very closely. It would be advantageous to be able to tune a similar transition in-situ. We will study the effect of different forms of light and will find that it is indeed possible to use longitudinal light to induce the same transition.

\section{Regularization of the undriven Hamiltonian}
\label{sec:regularize}
\begin{figure}
	\centering
	\includegraphics[width=1\linewidth]{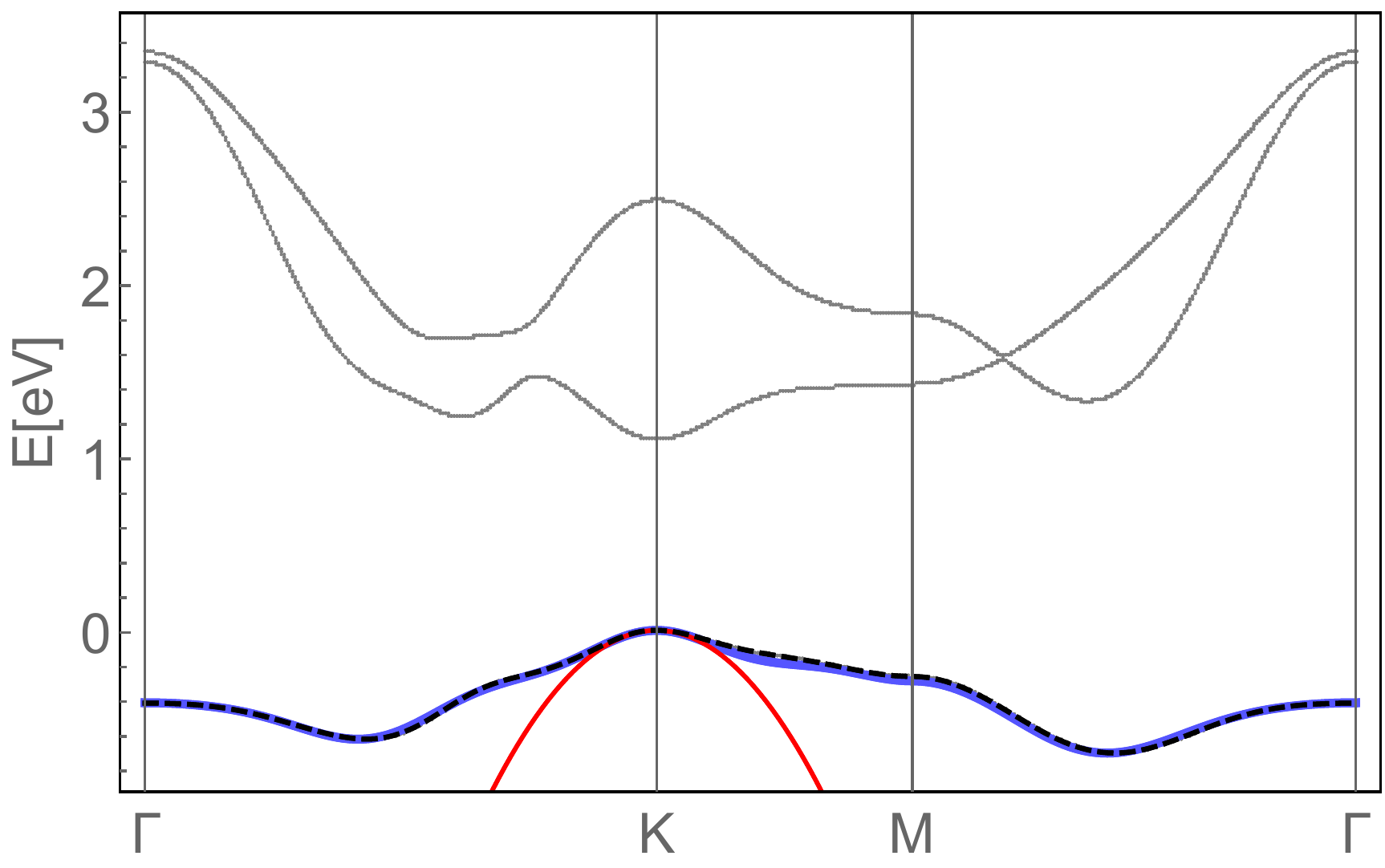}
	\caption{(Color online) Equilibrium band structure for $\text{MoTe}_2$  along a high-symmetry path in the BZ. Spin splitting effects are neglected \cite{liu2013three} especially since near the K point of interest the splitting is just a constant energy shift\cite{xiao2012coupled}. Shown in dashed black is the active low-energy band of $\text{MoTe}_2$ that enters in the approximately description of the effective Hamiltonian \eqref{Hd0} via $f(\vect k)$. In gray we show inactive bands that don't contribute to the description. In red we present the quadratic approximation used in Eq.\eqref{Hd0} and in blue the improved approximation. While we only plot a representative high symmetry path, the approximation is valid in the whole Brillouin zone.}
	\label{fig:betterbandstructapprox}
\end{figure}

After reviewing the equilibrium properties, we will investigate the properties of the system under the influence of different light sources. We first determine what frequency regimes are most likely to have an accurate description using effective Floquet Hamiltonians. First one should note that in order to be consistent with the low-energy effective Hamiltonian in Eq. \eqref{Hd0} we need to make sure that driving frequencies fulfill $\Omega < 1.1$ eV so as to avoid interband absorption to remote higher energy bands that are not captured with this Hamiltonian (see Fig.~\ref{fig:betterbandstructapprox}). 
We also neglect lower energy valence bands because near the $\vect K$ point they are spectrally isolated (the gap to additional valence bands is of the order $\sim 2$ eV \cite{pandey2020layer}).  This consideration does not apply to the the spin-orbit split partner of the top valence band, which is close in in energy ($\Delta_{SOC}\approx 220$meV \cite{Wu_2019}).  Our neglect of this band is, however, justified by the absence of optical coupling between the spin-orbit split partners at the top of the valence band.  In Appendix \ref{lightANDlimits} we provide some additional details. Precisely, this means that for frequencies close to $1.1$ eV we assume that the effects of light are dominated by single photon processes. Other bands that were neglected, such as ones coming from different orbitals, will also be ignored under the same assumption.

Before we study the effect of light, however, it is important to devise a relevant model that makes it easy to treat using high-frequency approximations in Floquet theory. Many of the estimates for such discussions rely on operator norms. The effective Hamiltonian in  Eq.\eqref{Hd0}, however, is not well-suited to such estimates because one of its constituents $f(\vect k)=-\frac{\vect k^2}{2 m^*}$ is unbounded. That the Hamiltonian is unbounded also leads to convergence issues in our numerics\footnote{Including arbitrarily many plane waves gives an arbitrary number of spurious high lying bands. These bands have Floquet copies associated with them that will intersect with actual bands. It is important to note that further bands can always be generated by including more Floquet copies and plane waves and these might also intersect. Therefore for such a model convergence for the quasi-energy band structure will not be reached.} and therefore it is best avoided by choosing a more well-suited effective model.

To sidestep these issues we realize that we need to offer a suitable replacement for $f(\vect k)$ that is bounded. The best choice is of course to make use of an improved approximation that captures some features on a more global scale. Such an expression is given below (the details of its derivation are given in Appendix \ref{better_single_layer_dispersion})
\begin{equation}
	\begin{aligned}
	\tilde f(\vect k-\vect K)\approx&C_0+\sum_{n=1}^5 C_n[2 \cos (n X_k) \cos (n Y_k)+\cos (2n X_k)]\\
	&+C_6[2 \cos (3 X_k) \cos (Y_k)+\cos (2 Y_k)]\\
	&+C_7[2 \cos (6 X_k) \cos (2 Y_k)+\cos (4 Y_k)]\\
	&+2 C_8 \left[\cos (8 X_k) \cos (2 Y_k)+\cos (7 X_k) \cos (3 Y_k)\right.\\
	&+\left.\cos (X_k) \cos (5 Y_k)\right]\\
	&+2 C_9 \left[\cos (9 X_k) \cos (Y_k)+\cos (6 X_k) \cos (4 Y_k)\right.\\
	&+\left.\cos (3 X_k) \cos (5 Y_k)\right]\\
	&+2 C_{10}[\cos (7 X_k) \cos (Y_k)+\cos (5 X_k) \cos (3 Y_k)\\
	&+\cos (2 X_k) \cos (4 Y_k)]\\
	&+2 C_{11}[\cos (5 X_k) \cos (Y_k)+\cos (4 X_k) \cos (2 Y_k)\\
	&+\cos (X_k) \cos (3 Y_k)]
	\end{aligned},
	\label{betterbandsEqs}
\end{equation}
where we introduced the shorthand notation $Y_k=(\sqrt{3}/2)a k_y $, $X_k=(1/2)ak_x$ and included a shift by $\vect K=(4\pi/3/a,0)$ to end up with a more convenient notation, where $a=3.472$\AA \ is the lattice constant of the unit cell \cite{liu2013three}. The coefficients in this expression are given in units of eV as
\begin{equation}
	\begin{aligned}
	&C_0=-0.4137;\quad C_1= -0.1046;\quad C_2= 0.0322\\
	&C_3= -0.0221;\quad C_4= -0.0080;\quad C_5= -0.0012\\
	&C_6= 0.0916;\quad C_7= 0.0060;\quad C_8= -0.0047\\
	&C_9= 0.0080;\quad C_{10}= -0.0055;\quad C_{11}= 0.0046
	\end{aligned}.
\end{equation}
To motivate this expression we recall that $f(\vect k)$ arises from one particular band in single layer $\text{MoTe}_2$ (see Appendix \ref{better_single_layer_dispersion}). The expression $f(\vect k)=-\frac{\vect k^2}{2 m^*}$ is a crude approximation of this band only valid near $\vect K$, and $\tilde f(\vect k)$  captures more features and is  better behaved in the sense that it is bounded, as shown in Fig.~\ref{fig:betterbandstructapprox}. Although we should still stress that it is most valid near $\vect K$ because some behaviour due to spin-orbit coupling away from the $\vect K$ point was neglected. We refer to Appendix \ref{better_single_layer_dispersion} for a more detailed discussion on the derivation of these results. We observe that the new approximation (blue) captures the band of interest (dashed black) more accurately than the quadratic approximation (red). Intralayer couplings between the blue band and the gray bands are assumed negligible because of the large energy gap. Since interlayer couplings are usually even weaker we assume that we can neglect the upper gray bands and other bands that are not displayed. 

The Hamiltonian now is the same as Eq.\eqref{Hd0}, just with the new $\tilde f$ instead of $f$, that is
\begin{equation}
\hspace*{-0.245cm}
\mathcal{H}_{\uparrow}=\begin{pmatrix}
\tilde f(R_{-\theta/2}(\vect k-\kappa_+)) +\Delta_{1} & \Delta_{T} \\
\Delta_{T}^{\dagger} & \tilde f(R_{\theta/2}(\vect k-\kappa_-)) +\Delta_{-1}
\end{pmatrix},
\label{new_eff_Ham_regularized}
\end{equation}
where all references to position dependence have been dropped to simplify the notation. A rotation matrix for momenta had to be introduced to measure momenta in both layers in the same coordinate system, which was necessary because $\tilde f(\vect k)$ is not rotationally symmetric unlike $f(\vect k)$. One should note that $\tilde f(\vect k)$ has the dominant bandwidth of the problem $\sim 0.7$ eV. We stress that our approximation means that only intra-valence band optical matrix elements are considered in this theory. We also stress that more realistic simulations that take other bands into account are needed but are beyond the scope of this work. Therefore, we continue by conventional estimates  \cite{Eckardt_2015,Vogl_2020,Bukov_2015} with our model, and find that driving frequencies $\Omega>0.7$ eV can be considered to be in a high-frequency limit from this theoretical point of view. We will neglect the influence that light in this frequency regime potentially has on phonons.

We also stress that the estimates for the high frequency regime window discussed here are specific to twisted MoTe$_2$ and that the high frequency regime for other TMDs may be different or may not even exist. However, the existence of a similar regime for other TMDs seems likely. Particularly, if we look at band structure plots shown in \cite{liu2013three} for different single  layer TMDs (single layer energies are the dominant energy scale and therefore serve well as first estimate), we find that each of the ones shown would likely permit a similar high frequency regime.

\section{Mathematical aspects of light-driven system}
\label{sec:coupl}
In this section we will discuss how different forms of light affect the Hamiltonian in Eq.\eqref{new_eff_Ham_regularized}. A periodic time-dependent Hamiltonian $H(t+T)=H(t)$ commutes with the generator of time translations, $e^{i(-iT\partial_t)}$. Therefore, similar to Bloch's theorem the wavefunction factorizes in the form $\Psi(t) =e^{-i\varepsilon t} u(t)$, where $u(t)$ is periodic in time and $\varepsilon$ is a constant called the quasi-energy \cite{PhysRevA.7.2203,Eckardt_2015}. Inserting this ansatz in the Schr\"odinger equation, one finds the Floquet-Schr\"odinger equation,
\begin{equation}
\varepsilon u(t)=\left[H(t)-i\partial_t \right]u(t),
\label{quasieneq}
\end{equation}
which now treats time on the same level as position--as an operator. If we expand Eq.\eqref{quasieneq} in terms of plane waves $e^{in\omega t}$, we find that the equation can be put in the form
\begin{align}
&\nonumber\begin{pmatrix}
\ddots&\vdots&\vdots&\vdots&\vdots&\vdots&\\
\cdots&H_1& H_0-\omega&H_{-1}&H_{-2}&H_{-3}&\cdots\\
\cdots&H_2&H_1&  H_0&H_{-1}&H_{-2}&\cdots\\
\cdots&H_3&H_2&H_1& H_0+\omega&H_{-1}&\cdots\\
&\vdots&\vdots&\vdots&\vdots&\vdots& \ddots
\end{pmatrix}\begin{pmatrix}
\vdots\\
u_{-1}\\
u_0\\
u_1\\
\vdots
\end{pmatrix}\\&=\varepsilon \begin{pmatrix}
\vdots\\
u_{-1}\\
u_0\\
u_1\\
\vdots
\end{pmatrix},
\label{quasi-en-eq}
\end{align}
where 
\begin{equation}
	 H_n=\frac{1}{T}\int_0^T dt e^{-i n\omega t} H(t)
\end{equation}
and 
\begin{equation}
	u_n=\frac{1}{T}\int_0^T dt e^{-i n\omega t}.
\end{equation}
This representation is exact, and suitable for numerical implementations upon truncation to finite $|n|\leq n_{\mathrm{max}}$.  Physical considerations must be made to choose a reasonable $n_{\mathrm{max}}$ and convergence can be checked by changing this value and seeing that the results around $n=0$ are numerically unchanged.  In the next sections, we use this representation to obtain our results. 

\subsection{Circularly polarized light}

\begin{figure}[H]
	\centering
	\includegraphics[width=0.9\linewidth]{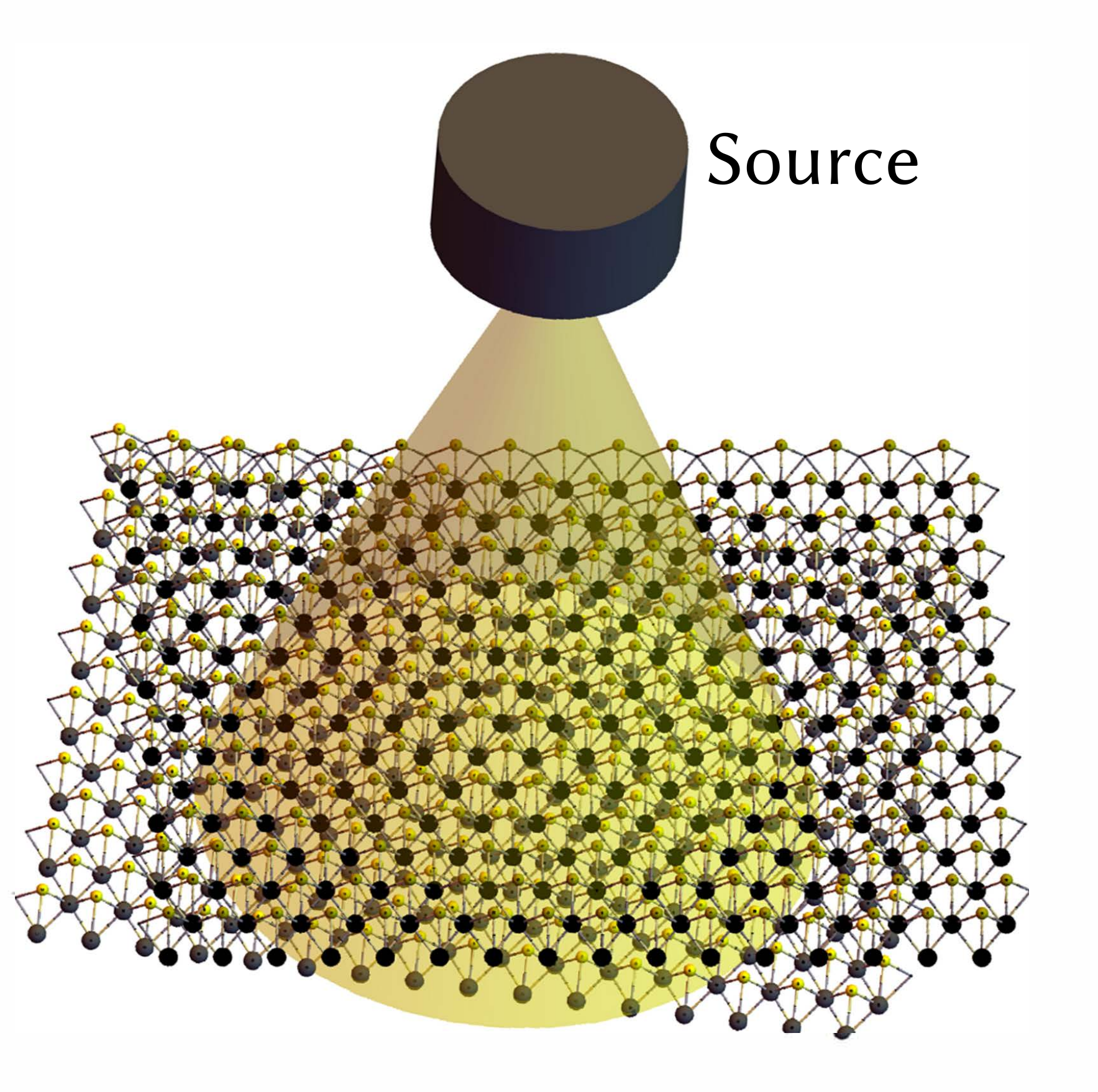}
	\caption{(Color online) Cartoon of a twisted TMD subjected to circularly polarized light from a distant source in free space..}
	\label{fig:circpollightfigure}
\end{figure}

We first consider the twisted TMD when subjected to circularly polarized light as shown in Fig. \ref{fig:circpollightfigure}. Circularly polarized light is described by a vector potential that has components only in the plane and can be introduced very simply via the usual minimal substitution $\vect k\to \vect k-\vect A$ and $\vect A=(A_x\cos(\Omega t),A_y\sin(\Omega t))$. Here we neglect the effect light has on interlayer hopping elements because i)  interlayer hopping is dominated by hopping processes that are almost perfectly in z-direction  (in the Peirls substitution this means that for circularly polarized light $\vect Ad\vect l\approx 0$, where $d\vect l$ is the displacement element between atoms in adjacent layers) and ii) interlayer hoppings are initially so small compared to intra-layer hoppings that corrections to them can be neglected when compared to corrections due to the intra-layer hoppings.

One may now want to compute the components $H_n$ that are needed for a numerical implementation. The result is given below

\begin{equation}
\mathcal{H}_{n,\uparrow}^{\mathrm{circ}}( \vect r)=\tilde H_n^\mathrm{circ}+\Delta_V(\vect r)\delta_{0n},
\label{Hnsforcircpol_light}
\end{equation}
where
\begin{equation}
\Delta_V(\vect r)=\begin{pmatrix}
\Delta_{1}(\vect r) & \Delta_{T}(\vect r) \\
\Delta_{T}^{\dagger}(\vect r) & \Delta_{-1}(\vect r)
\end{pmatrix}
\end{equation}
with $\Delta_{T}(\vect r)$ given by Eq.\eqref{Tunneling}, $\Delta_{\pm1}(\vect r)$ given by Eq.\eqref{delta_l}, and 
\begin{equation}
	\tilde H_n^{\mathrm{circ}}=\begin{pmatrix}
	\tilde f_n(R_{-\theta/2}(\vect k-\kappa_+)) & 0\\0& \tilde f_n(R_{\theta/2}(\vect k-\kappa_-))
	\end{pmatrix},
	\label{HnCircPol}
\end{equation}
where the functions $\tilde f_n(\vect k)$ can be found by first decomposing $\tilde f$ into a sum of $q_{n_1,n_2}^k=\cos(n_1 X_k+n_2Y_k)$. Afterwards one may apply minimal coupling to the vector potential $k\to k-A(t)$ and finds that the time dependence in the resulting Hamiltonian only enters through $q_{n_1,n_2}^{k-A(t)}$. One may therefore compute the integrals $\tilde q^n_{n_1,n_2}=\frac{1}{T}\int_0^T dt q^{k-A(t)}_{n_1,n_2}e^{-in\omega t}$. The result of this analysis allows us to find $\tilde f_n$. One merely has to replace every term $q_{n_1,n_2}$ in $\tilde f$ by
\begin{equation}
\begin{aligned}
	&\tilde q^n_{n_1,n_2}=e^{i m \left(\tau (n_1,n_2)+\frac{\pi }{2}\right)} J_m\left(\frac{A a_0}{2}  \sqrt{n_1^2+3 n_2^2}\right) \times\\
	&\times\cos \left(X_k n_1+Y_k n_2+\pi  \left(\frac{m}{2}+\frac{2 n_1}{3}\right)\right)\\
	&-\cos \left(\frac{2 \pi  n_1}{3}\right) \delta _{0,m}
	\end{aligned},
\end{equation}
where $J_m$ is the $m$-th Bessel function of the first kind. Additionally we have 
\begin{equation}
	\tau_{n_1,n_2}=\begin{cases}
	\theta+\frac{\pi}{2};& n_2=0\\
	 \tan ^{-1}\left(\frac{n_1}{\sqrt{3} n_2}\right)+\frac{\pi}{2}(1-  \mathrm{sgn}(n_2)); &\text{else}.
	\end{cases}
\end{equation}

We have now derived the mathematical formulation in the extended space picture for circularly polarized light. Before we discuss the physical consequences of the light, let us first turn to the mathematical formulation for light from a waveguide.

\subsection{Light from a waveguide}

\begin{figure}[H]
	\centering
	\includegraphics[width=1\linewidth]{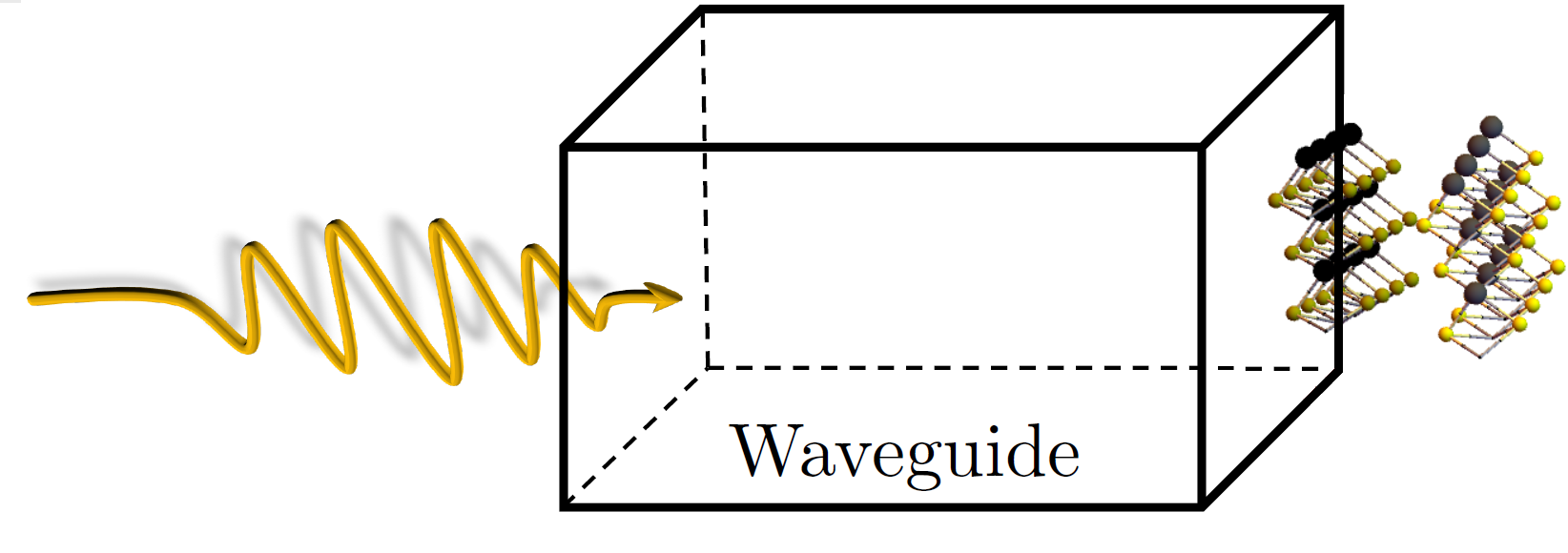}
	\caption{(Color online) Cartoon of a twisted TMD double layer subjected to a transverse magnetic modes from a waveguide.}
	\label{fig:waveguidefigure}
\end{figure}

Let us next consider the transverse magnetic mode, which has an electric field component in the interlayer direction. Light subjected to a waveguide may have such a mode as displayed in Fig. \ref{fig:waveguidefigure}.
In a finite region of space it may accurately be approximated by the vector potential $\vect A=A\cos(\Omega t)\hat z$, which has only a longitudinal component  \cite{Vogl_2020interlayer,rodriguezvega2020floquet}. This form of light can be included in the Hamiltonian via the replacement $w\to e^{-iAa_L\cos(\Omega t)}w$ because it only affects interlayer couplings through a Peierls substituion. This up to a gauge transformation is equivalent to an electric potential between layers. More details can be found in Appendix \ref{lightANDlimits}.  We assumed that the thickness of the twisted TMD bilayer $a_L\approx 7.1\AA$ \cite{wang2019weakened} is much smaller than the wavelength $\lambda$ of light $a_L\ll \lambda$.

The relevant quantities $H_n$ for an extended space numerical implementation are found very conveniently as,
\begin{equation}
	\mathcal{H}_{n,\uparrow}^{\mathrm{circ}}( \vect r)=\tilde H\delta_{n,0}+\tilde \Delta_{V,n}(\vect r),
	\label{waveguideHn}
\end{equation} 
where (with position dependence dropped to simplify the notation)
\begin{equation}
	\hspace*{-0.245cm}
	\tilde H=\begin{pmatrix}
	\tilde f(R_{-\theta/2}(\vect k-\kappa_+)) +\Delta_{1} & 0 \\
	0 & \tilde f(R_{\theta/2}(\vect k-\kappa_-)) +\Delta_{-1}
	\end{pmatrix}
\end{equation}
and
\begin{equation}
	\hspace*{-0.245cm}
	\tilde \Delta_{V,n}(\vect r)=\begin{pmatrix}
	0 & \Delta_{T,n}(\vect r) \\
	\Delta_{T,n}^{\dagger}(\vect r) & 0
	\end{pmatrix},
\end{equation}
where
\begin{equation}
	\Delta_{T,n}(\vect r)=i^nJ_n(a_L A)\Delta_T(\vect r).
\end{equation}
Here $J_n$ is the $n$-th Bessel function of the first kind.

\section{Physical aspects of light-driven twisted TMD bilayers}
\label{sec:phys_imp_liights}
Now that we have collected all the necessary parts for a numerical implementation of the light-matter coupling, we discuss the impact that different forms of light have on the tTMD.

\subsection{Effect of circularly polarized light}
We consider the high-frequency regime and use a van Vleck expansion to first order,
\begin{equation}
	H_{vV}\approx H_0+\sum_n\frac{[H_{-n},H_{n}]}{2n\Omega}.
\end{equation}
We recognize immediately from Eq.\eqref{HnCircPol} that $[H_{-n},H_{n}]=0$ for $n\neq 0$. Therefore, the first order corrections, those that goes as $\Omega^{-1}$, vanish. For the zeroth order contribution we will focus on the low-energy results that for small angles correspond to small momentum displacements from the $\vect K $ point (not quasi-momenta) and we consider relatively weak drives $A$. Thus, we are allowed to do a Taylor expansion around small momenta and $A$. 

The result one finds is quite lucid. It is only a constant quasi-energy shift $-\frac{A^2}{2m^*}$. This is also confirmed numerically: for small angles, low energies and in the high-frequency regime the only relevant change appears to be this shift. This is in stark contrast to twisted bilayer graphene where circularly polarized light has a profound effect on the spectrum. For graphene the lowest order effect is the opening of a gap; for the twisted TMD there already is a significant gap and therefore the effect is small. The results presented for far are only valid in the extreme high-frequency regime. As the frequency decreases and the Floquet zones overlap, other effects can be expected. Instead of pursuing this route, we consider different forms of light, namely longitudinal light, which can have a profound effect on the system at leading order.


\subsection{Effect of light from a waveguide}
We will consider only the high-frequency regime for simplicity and make use of a van Vleck expansion to first order. Similar to circularly polarized light, the first order contribution vanishes because $[H_{-n},H_{n}]=0$ for $n\neq 0$ as one can see from Eq.\eqref{waveguideHn}. The effect of the zeroth order contribution is just a renormalization of the interlayer coupling strengths. If we look at this in terms of a skyrmion pseudo-spin lattice derived from $\vect \Delta (\vect r)$, then we find that the texture essentially remains the same, aside from the length of vectors in the $x$-$y$ plane are shortened by a factor of $J_0(a_LA)$, the $z$-direction remains unchanged. Such textures are a notable indicator of possible topological properties, which must be tested by evaluating candidate topological invariants.

\subsubsection{Effect on band topology and spectrum}
\begin{figure}[t]
	\centering
	\includegraphics[width=1\linewidth]{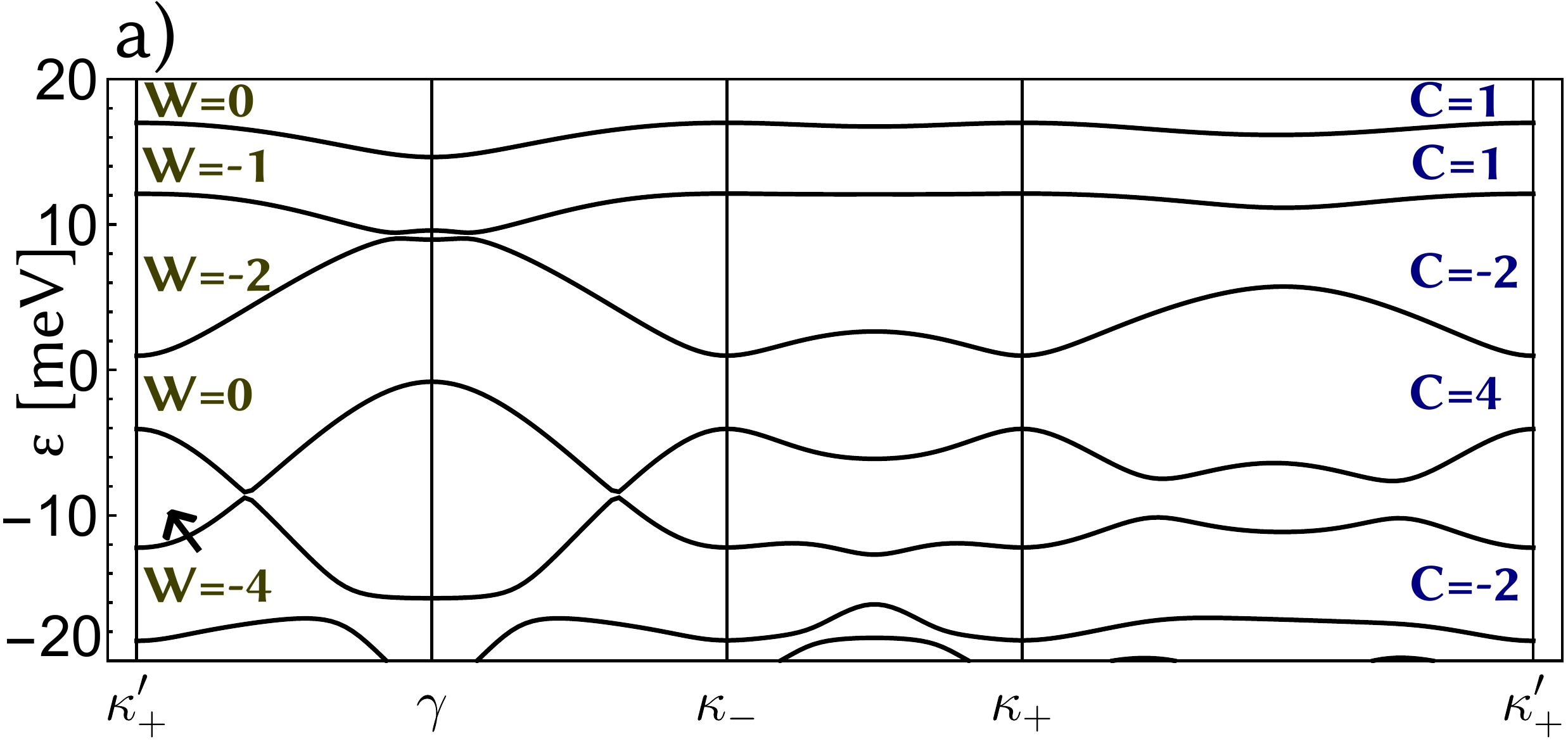}
	\includegraphics[width=1\linewidth]{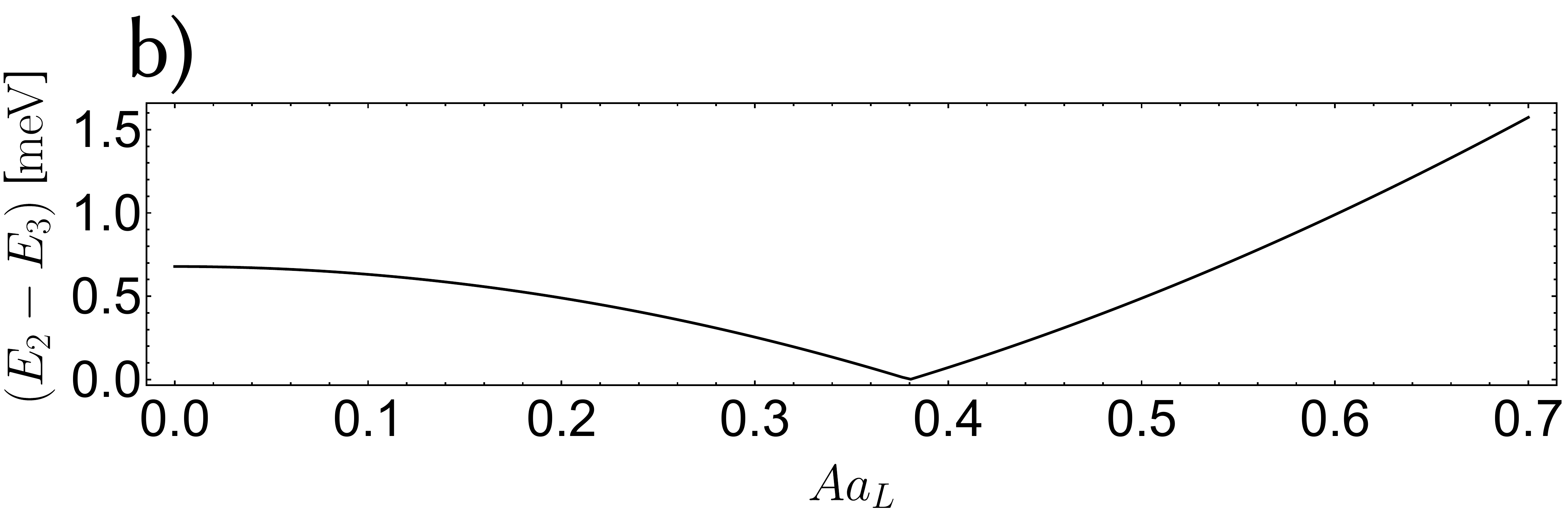}
	\includegraphics[width=1\linewidth]{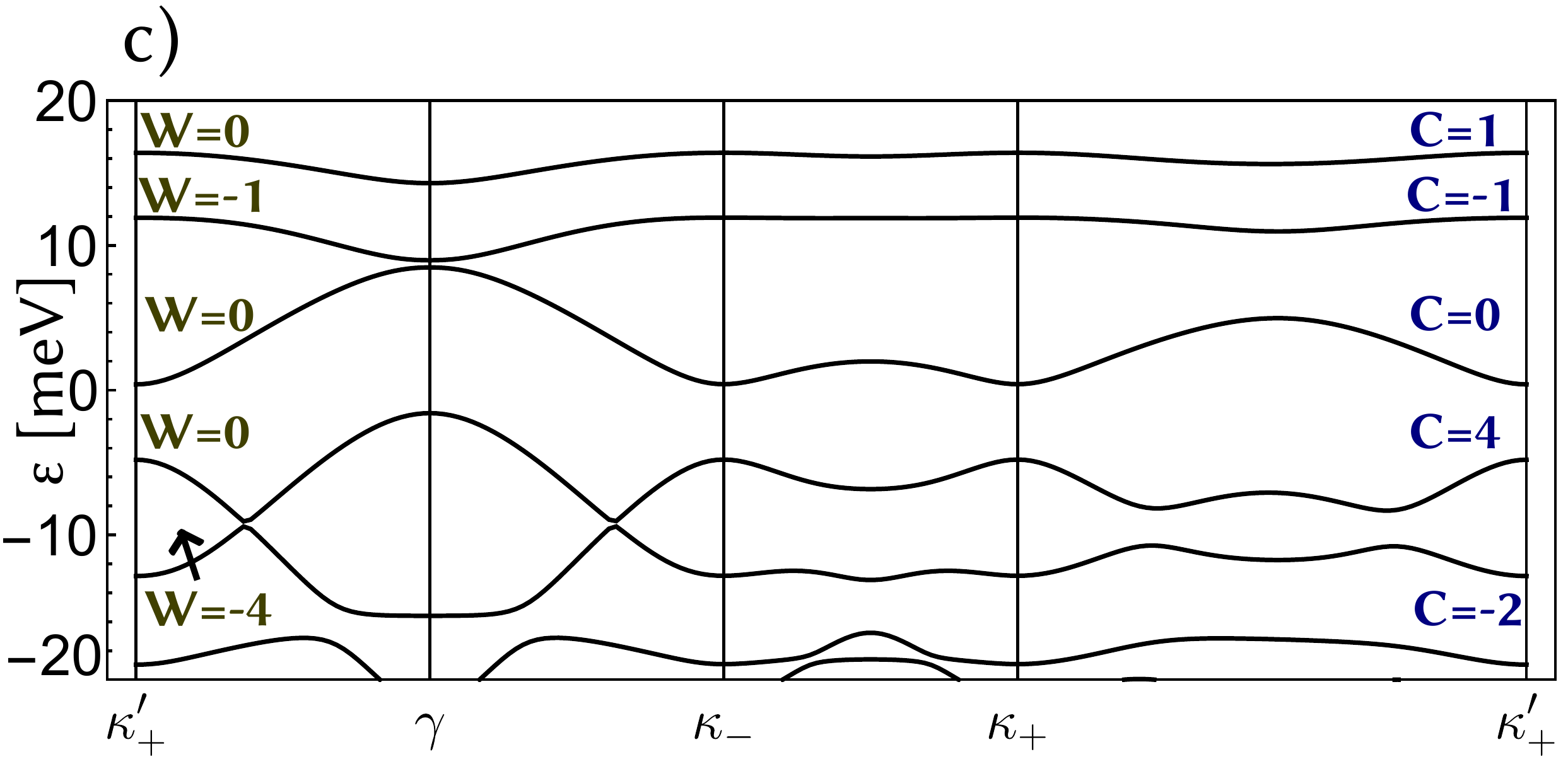}
	\caption{The figure shows that one can use light from a waveguide to close a gap between the second and third band (measured from the ``top") of the spectrum at a twist angle of $1.96^\circ$. In subfigure a) we show the system subjected to waveguide light of frequency $0.7$eV and driving strength $Aa_L=0.1$, in subfigure b) we show that the gap between band two and three at the $\gamma$ point eventually closes and reopens as a function of driving strength $Aa_L$. Subfigure c) shows the bands when subjected to driving strengths $Aa_L=0.5$ with the gap reopened. The physical realizability of this value for $Aa_L$ is discussed at the end of section \ref{sec:spintexture} We included blue insets with the Chern numbers $C$ and winding numbers $W$. The computations were made by including three Floquet copies and are converged. }
	\label{fig:gapclosesbydrive}
\end{figure}

\begin{figure}[]
	\centering
	\includegraphics[width=1\linewidth]{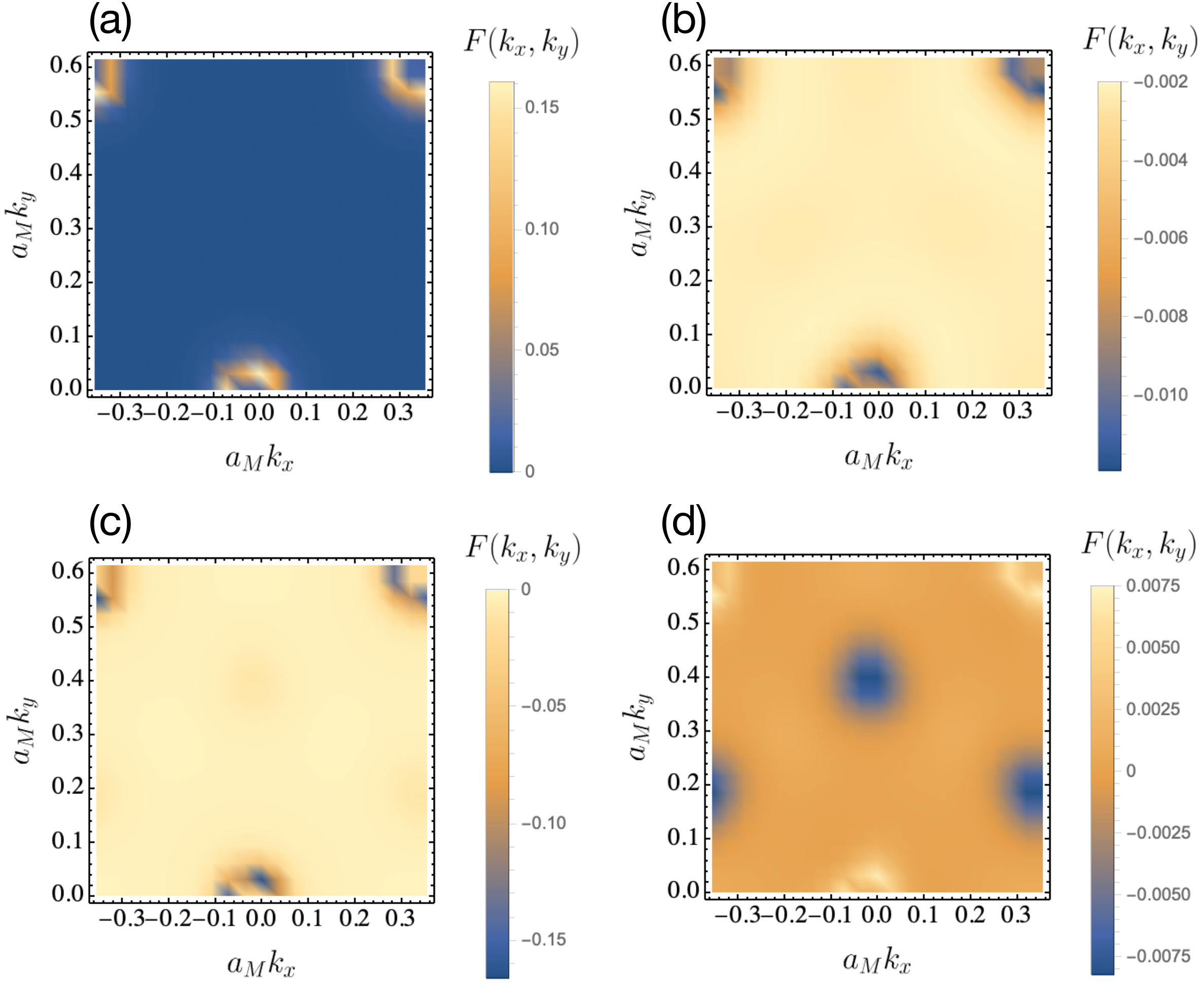}
	\caption{ (Color online) Floquet Berry curvature for the four top Floquet quasienergies in the first Floquet zone before (a-c) and after the gap closing (b-d) for the same band. The corresponding band Chern numbers are (a) $C_2 = 1$ ($a_M A = 0.1$), (b) $C_2 = -1$ ($a_M A = 0.5$), (c) $C_3 = -2$ ($a_M A = 0.1$), and (d) $C_3 =0$ ($a_M A = 0.5$). Here $\theta = 1.95^\circ$ and the driving frequency $\Omega=0.7$eV. }
	\label{fig:berry_cruv_1}
\end{figure}

In the regime of interest, the effective Floquet Hamiltonian to zeroth order in the van Vleck expansion is given by (first order corrections vanish) 
\begin{equation}
\hspace*{-0.245cm}
\mathcal{H}_{\uparrow}=\begin{pmatrix}
\tilde f(R_{-\theta/2}(\vect k-\kappa_+)) +\Delta_{1} & J_0(a_LA)\Delta_{T} \\
J_0(a_LA)\Delta_{T}^{\dagger} & \tilde f(R_{\theta/2}(\vect k-\kappa_-)) +\Delta_{-1}
\end{pmatrix},
	\label{Hdwaveguide}
\end{equation}
where the position dependence of $\Delta_i$ was dropped for a shorter notation. It is important to stress that this Hamiltonian is exact in the infinite-frequency regime. From the structure of the effective Hamiltonian, we find that light from a waveguide decreases the strength of the interlayer coupling, leading to an effective change in the twist angle. Particularly, by tuning the properties of the laser, we can induce quasienergy band closings and a subsequent band opening when one is close to but above to the critical angle $\theta^*\approx 1.85^\circ$, as shown in Fig. \ref{fig:gapclosesbydrive}. This leads to a change in the band Chern numbers. In Fig. \ref{fig:berry_cruv_1} we plot the Floquet Berry curvature for $a_M A = 0.1$ (a-c) and $a_M A = 0.5$ (b-d). 

In a time-independent system the bulk-edge correspondence tell us that Chern numbers can be used to compute the number of edge states between two adjacent materials. However, in a periodically driven system this bulk-edge correspondence is  not determined by the Chern number but by the winding number $\mathcal{W}[\mathcal{U}_\varepsilon]$. Unlike the Chern number that is associated with a band, the winding number is associated with a gap at quasi-energy $\varepsilon$. The winding number for a two-dimensional periodically driven system is given as
\begin{equation}
\mathcal{W}[\mathcal{U}]=\frac{1}{8 \pi^{2}} \int d t d \boldsymbol{k} \operatorname{Tr}\left(\mathcal{U}^{-1} \partial_{t} \mathcal{U}\left[\mathcal{U}^{-1} \partial_{k_{x}} \mathcal{U}, \mathcal{U}^{-1} \partial_{k_{y}} \mathcal{U}\right]\right),
\end{equation}
and $\mathcal{U}_{\varepsilon}$ is a modified time evolution operator \cite{PhysRevX.3.031005}. Similar to Chern numbers we also find that the winding number that is associated with the gap between the second and third quasienergy band changes. This indicates a change in number of edge states in a sample with open boundary conditions (corresponding to a finite material with physical boundaries).

\subsubsection{Effect on the pseudo-spin texture}
\label{sec:spintexture}
\begin{figure}[t]
	\centering
	\includegraphics[width=1\linewidth]{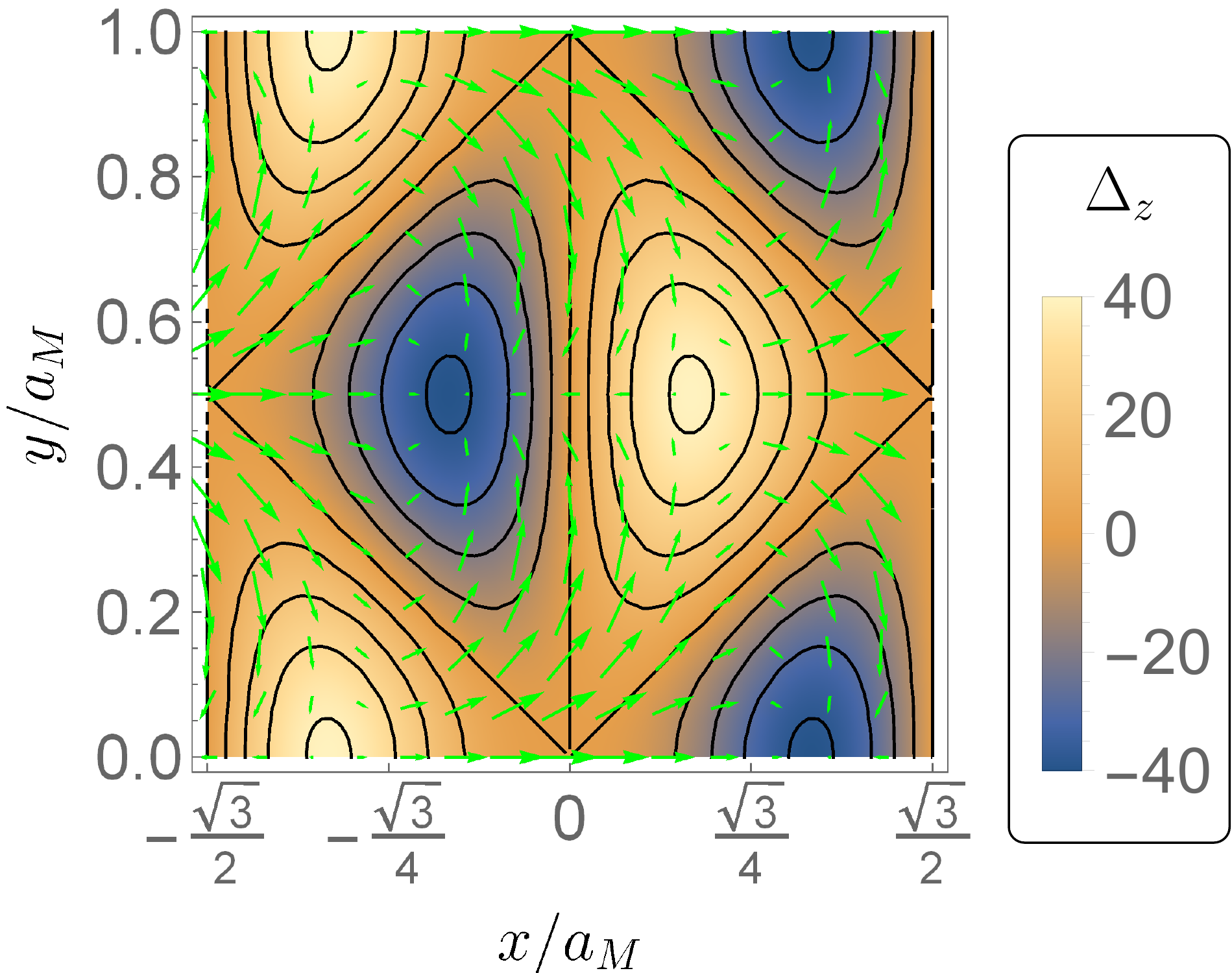}
	\caption{(Color online) Plot of $\vect \Delta(\vect r)$ for negative $J_0(Aa_L)$, where $J_0(Aa_L)\Delta_{x,y}$ correspond to the green arrows and the density plot corresponds to $\Delta_z$.}
	\label{fig:skyrmions2}
\end{figure}

In addition to the band topology we also find that light has an impact on the pseudo-spin texture. Specifically, we find that the terms
\begin{equation}
	\Delta_{x,y}\to J_0(Aa_L)\Delta_{x,y}
\end{equation}
are renormalized. This of course at first does not impact the structure seen in Fig.\ref{fig:skyrmions} since only their magnitude changes unless $J_0$ changes sign. In the latter case the arrows in the $x$-$y$ plane change direction. This is shown in Fig. \ref{fig:skyrmions2}.  Notice that the green arrows in Fig. \ref{fig:skyrmions2} point in the opposite direction compared to the red arrows in Fig. \ref{fig:skyrmions}.  In our case this directional change occurs at $Aa_L= j_{0,1}\approx 2.405$, where $j_{n,1}$ is the first zero of the nth Bessel function of the first kind. 

Interestingly, this observation is not an artifact of the approximation, which is expected to work for $Aa_L<1$.  We have found the same result by exact numerical calculations. It should also noted that this value for $Aa_L$, while large, is just within experimental reach: for $\Omega=0.7$ eV and an electric field strength $E=25$MV/cm, we find that $Aa_L\approx 2.5$.

\begin{figure}[]
	\centering
	\includegraphics[width=1\linewidth]{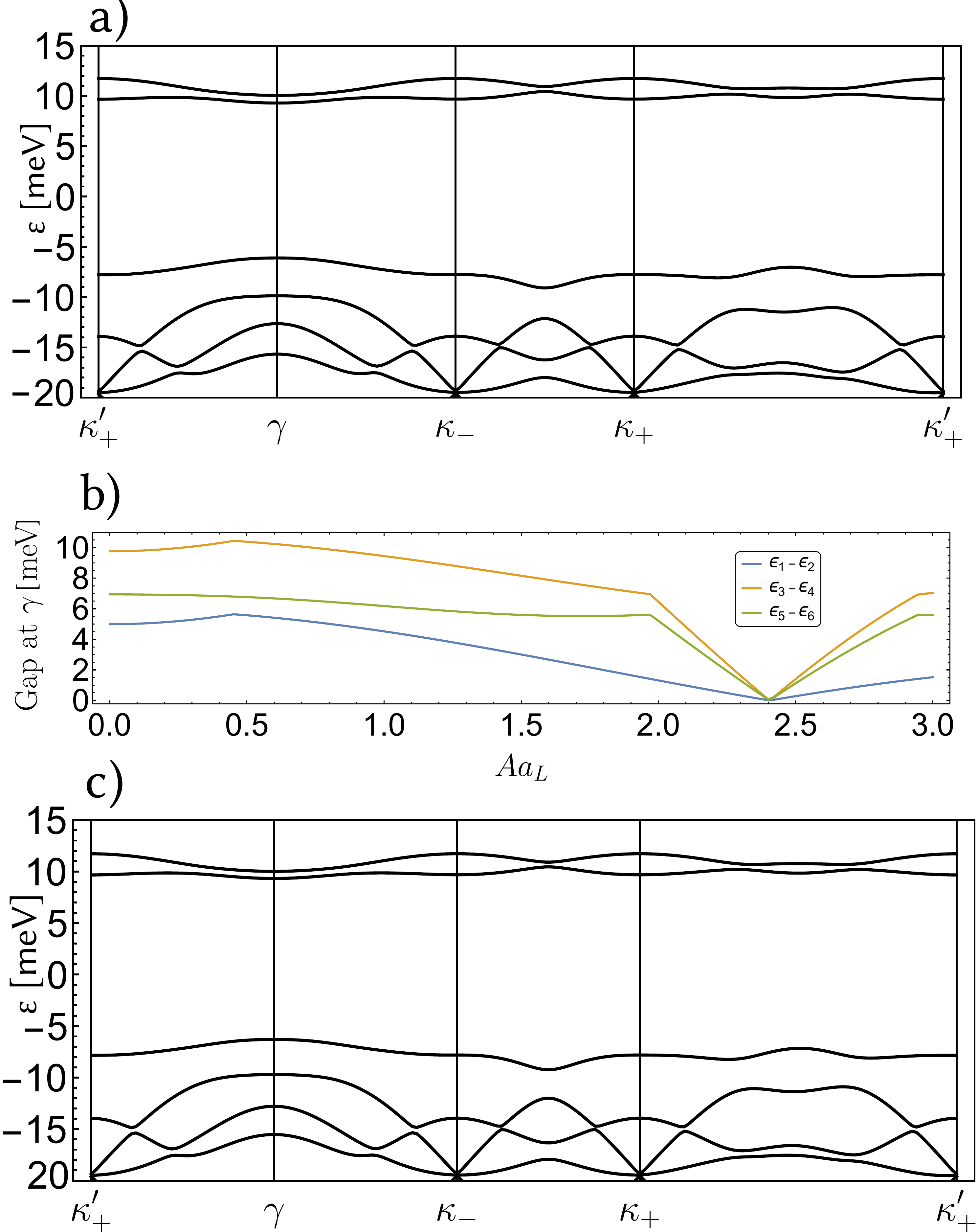}
	\caption{(Color online) The figure illustrates that for strong drives at an angle of $\theta = 2^\circ$ gaps for many bands the gap at the $\gamma$ point closes and reopens. In subfigure a) we show a plot of the quasi energy band structure along a high symmetry path for $Aa_L=0.9j_{0,1}$, in subfigure c) for $Aa_L=1.1j_{0,1}$. In subfigure b) we plot several band gaps at the $\gamma$ point as a function of $Aa_L$.
	The results were computed numerically using 3 Floquet copies and a driving frequency $\Omega=0.7$eV. }
	\label{fig:strongdriventh20p9j01}
\end{figure}

Now one can imagine that if something drastic happens in real space like in our case a sudden change in the pseudo-spin texture, there may also be something drastic associated with it in momentum space such as changes in the quasienergy band structure. Indeed, we find that in the infinite-frequency limit of Hamiltonian \eqref{Hdwaveguide} multiple band gaps close at this point and reopen afterwards. For the finite but large frequency limit seen in Fig. \ref{fig:strongdriventh20p9j01} we find that this corresponds to band gaps that almost close with gaps as small as $10^{-6}$ eV for a driving frequency of $\Omega=0.7$ eV. 

\begin{figure}[]
	\centering
	\includegraphics[width=1\linewidth]{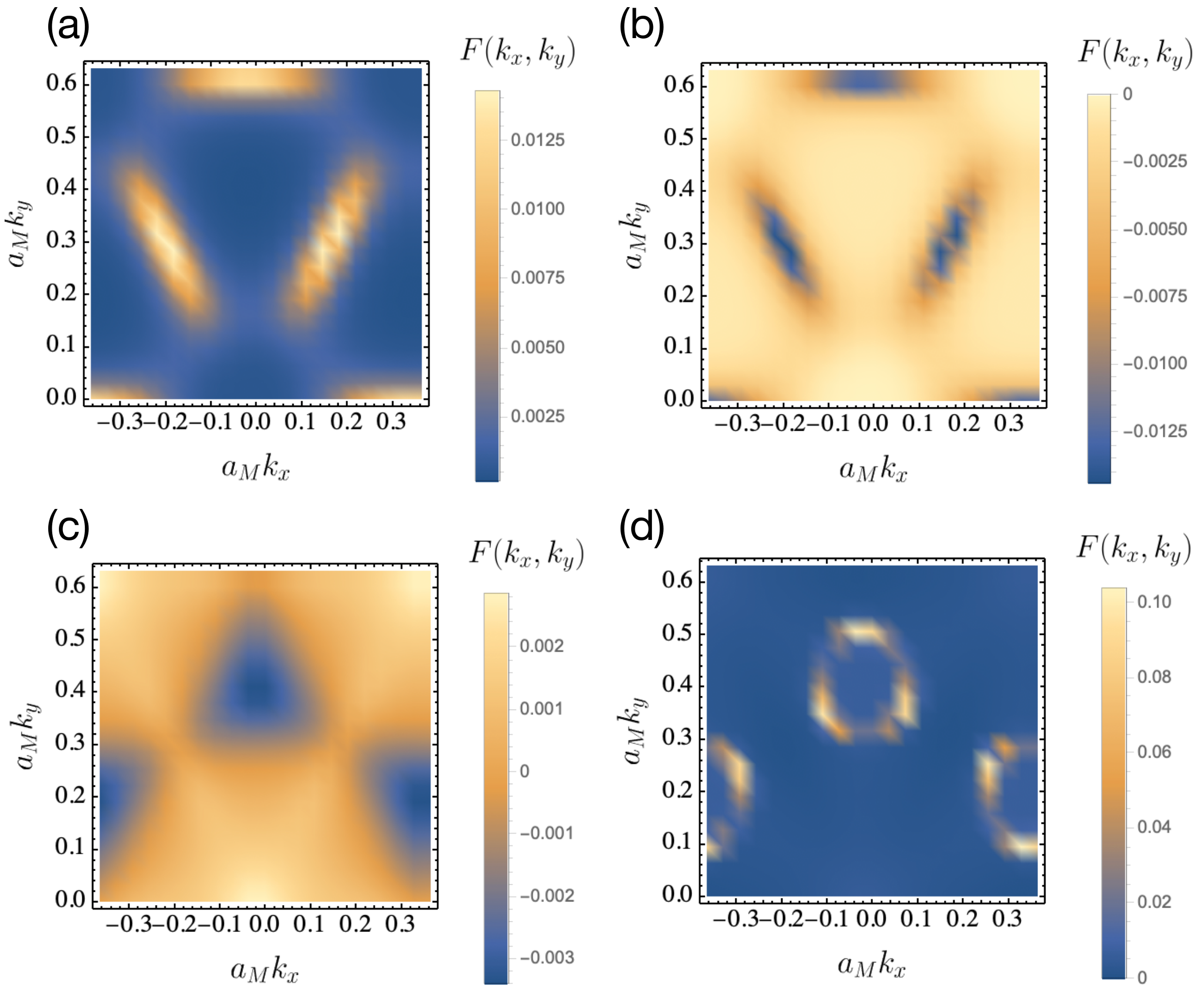}
	\caption{(Color online) Floquet Berry curvature for the four top Floquet quasienergies in the first Floquet zone. The corresponding band Chern numbers are (a) $C_1 = 1$, (b) $C_2 = -1$, (c) $C_3 = 0$, and (d) $C_4 =2$. The parameters are $\theta = 2^\circ$, $Aa_L=0.9j_{0,1}$, and driving frequency $\Omega=0.7$eV. }
	\label{fig:berry_cruv_2}
\end{figure}

In Fig \ref{fig:berry_cruv_2} we show the Floquet Berry curvature $F$ for the top four quasienergy bands in the first Floquet zone for $\theta = 2^\circ$ and $Aa_L=0.9j_{0,1}$.  Interestingly, even in the infinite-frequency regime the band gap closings are not associated with a change in Chern and winding numbers so there is no topological transition. The kinks in the plot at around $Aa_L\approx0.4$, $Aa_L\approx 1.96$ and $Aa_L\approx 2.95$ are associated with a band gap closing between band two and three near angle $\theta^*$. The other two kinks at $Aa_L\approx 1.96$ and $Aa_L\approx 2.95$ are associated with a band gap that closes between the fourth and fifth band. 

\section{Conclusions}
\label{sec:conculsions}
In conclusion, we have investigated the effect of circularly and longitudinally polarized light on twisted transition metal dichalcogenides (tTMD) and found that the effect of circularly polarized light in the high-frequency regime leads to a trivial shift in quasi-energy spectrum, in stark contrast to the result for twisted bilayer graphene. However, longitudinal light emanating from a waveguide can directly renormalize the interlayer tunneling amplitude and lead to topological transitions. We have also computed the Floquet Berry curvature for these systems to better understand the connection to the winding number, ${\cal W}$, changes indicating the topological transitions.  In addition, inter-layer pseudo-spin skyrmion texture manipulation in experimentally accessible regimes is possible. Experimental signatures of the light-induced topological transition could be detected in optical conductivity measurements\cite{mciver2020light}.

Similar effects could be expected in systems with similarly weak interlayer tunneling.  Our results show dramatically that tTMDs bring new physics to the table in the non-equilibrium regime beyond that of twisted bilayer graphene.   An exciting frontier to consider in future work is the role of phonons (that may be selectively excited by light) on the electronic and magnetic properties of tTMDS. We hope this work will help motivate further non-equilibrium studies of tTMDs to further explore the rich scope of possible behavior in this material class. 


\section{Acknowledgements}

This research was primarily supported by the National
Science Foundation through the Center for Dynamics and
Control of Materials: an NSF MRSEC under Cooperative Agreement No. DMR-1720595 and partial support from grant DMR-1949701.  A. H. MacDonald acknowledges support from Welch grant Welch TBF1473.

\bibliography{literature}

\appendix

\section{Origin of model and improvements}
\subsection{Origin of the model and couplings to circularly and longitudinally polarized light}
\label{lightANDlimits}
Because we are dealing with an effective model it is worthwhile to understand its origin. This will help to see how the light couples and frequency regime for which the description is valid.

Let us first consider a single layer of a TMD. For this case the Hamiltonian can be approximated  \cite{xiao2012coupled,liu2013three} as
\begin{equation}
\begin{aligned}
h^{(1)}_{\vect k}=&a_0t(k_x\sigma_x-ik_y\sigma_y)\otimes\mathbb{1}_2^s+\frac{\tilde\Delta_1}{2}(\sigma_z+\mathbb{1}_2^p)\otimes \mathbb{1}_2^s\\
&+\frac{\tilde\Delta_{-1}}{2}(\mathbb{1}_2^p-\sigma_z)\otimes \mathbb{1}_2^s-\lambda \frac{\sigma_z-\mathbb{1}_2^p}{2}\otimes \tau_z.
\end{aligned}
\label{kp1}
\end{equation}
This model can be derived from  a nearest neighbor tight binding model of a triangular lattice that involves the d-orbital combinations $d_{z^2}$ and $1/\sqrt{2}(d_{x^2-y^2}+id_{xy})$ of $Mo$. These orbitals dominate the bandstructure for energies close to the Fermi energy and near the $\vect K$ points \cite{liu2013three}. Other types of hopping such as indirect hopping processes e.g. via $Te$ atoms or between other orbitals are neglected. In keeping less orbitals we also lose some bands. We will assume that the bands we dropped do not couple to bands we kept despite some of them being close in energy, i.e. we will assume that the Hamiltonian is block-diagonal and the bands we keep are in a block separate from the rest of the Hamiltonian. The parameter $\lambda$ appears due to on-site spin-orbit couplings  \cite{xiao2012coupled,liu2013three}.  Couplings between different $\vect K$ points are neglected and the switch between inequivalent $\vect K\to\vect K^\prime$ can be achieved by $k_x\to-k_x$ and $\lambda\to-\lambda$ \cite{xiao2012coupled,liu2013three}.

One should note that $\tau_i$ are the Pauli matrices and $\mathbb{1}^s_2$  the identity matrix that are acting in spin space. The $\sigma_i$ and $\mathbb{1}_2^p$ are the corresponding Pauli matrices and the identity that are acting in  a pseudo-spin space spanned by pseudo-spin states $d_{z^2}$ and $1/\sqrt{2}(d_{x^2-y^2}+id_{xy})$ \cite{liu2013three}.

If there is no interaction between layers an additional layer can be added by just having two copies of the Hamiltonian
\begin{equation}
h_{\vect k}^{(2)}=h^{(1)}_{\vect k}\otimes \mathbb{1}_2^l,
\label{kp2}
\end{equation}
where $\mathbb{1}_2^l$ is the identity matrix acting in layer space. Consistent with the intra-atomic approximation for spin orbit couplings we assume that the coupling between layers does not mix spins. Therefore following \cite{Rost_2019} we find that interlayer couplings can be treated in the form
\begin{equation}
H_{inter}=T(\vect d_0)\otimes \mathbb{1}_2^s\otimes \frac{\gamma_x+i\gamma_y}{2}+T^\dag(\vect d_0)\otimes \mathbb{1}_2^s\otimes \frac{\gamma_x-i\gamma_y}{2},
\label{kp3}
\end{equation}
where $\gamma_i$ are Pauli matrices in layer space and $\vect d_0$ is a displacement between layers. However, the interaction between the layers also modifies  $\tilde\Delta_i\to \tilde\Delta_i(\vect d_0)$ because each layer of $\text{MoTe}_2$ is not planar with $Te$ atoms protruding - the atoms get different energy contributions from adjacent layers.

We stress that this form can be derived from a tight binding model and that the couplings in $T(\vect d_0)$ are proportional to interlayer hoppings in a tight binding model. The twisted case can be described by replacing $\vect d_0\to \theta \hat z\times \vect r$  \cite{Wu_2019} and shifting the momenta of the momentum operator of the upper layer by $\kappa_-$ and the one of the lower layer by $\kappa_+$ to account for the shift of $\vect K$ points due to the rotation.

Since we now fully understand how this model is related to a tight binding description it is now easy to see how light of different types that is described by a vector potential $\vect A$ can be incorporated via a Peierls substitution $t_{\vect R\vect R^\prime}\to e^{-i\int_{\vect R}^{\vect R^\prime} \vect Ad\vect r}t_{\vect R\vect R^\prime}$.  Let us first consider the case $\vect A=(A_x,A_y,0)$ - like in elliptically polarized light. The approach results in the replacement $k_i\to k_i-A_i$. This is sufficient if we assume that interlayer hopping is dominated by processes where two atomic sites that interact are almost on top of each other because we then have negligible displacements in the $x-y$ direction. In this case $\int_{\vect R}^{\vect R^\prime} \vect Ad\vect r$ is small for the interlayer couplings in $T(\vect r)$ and the effect can be neglected. Next we consider the case $\vect A=(0,0,A_z)$,  found in light coming from a waveguide \cite{Vogl_2020interlayer}. For simplicity we will assume that the distance between the two layers is approximately a constant $a_L$. Then we find that the Peirls substitution results in a replacement $T(\vect r)\to e^{-iAa_L}T(\vect r)$.

Now let us relate this model to the one in equation \eqref{Hd0}. We first see that the Hamiltonian is diagonal in spin space and it is therefore valid without making a further approximation to only consider one spin species since there is no couplings. Even in the driven case for effective Hamiltonians this is valid because commutators that appear in an effective Hamiltonian do not break a block-diagonal structure (e.g. in the Magnus expansion $H_{\mathrm{eff}}\approx \frac{1}{T}\int_0^TH(t)+\frac{i}{2T}\int dt\int dt_1[H(t_1),H(t)]$) .

Therefore we will focus on spin $\uparrow$. We see that the Hamiltonian has the same form as in the appendix of \cite{Wu_2019}  but with the following values
\begin{equation}
\tilde \Delta_1=\Delta_g+\Delta_1;\quad \tilde\Delta_{-1}=\Delta_{-1},
\end{equation}
where $\Delta_g$ is a large gap of $\sim 1eV$ and $\Delta_i$ is given in equation \ref{delta_l}.
Furthermore we have
\begin{equation}
\begin{aligned}
T(\vect d_0)&=\begin{pmatrix} 
w_2 & w_{3} \\
w_{3} & w
\end{pmatrix}\\
&+
\begin{pmatrix} 
w_2 & w_{3} e^{-i2\pi/3} \\
w_{3} e^{i2\pi/3} & w
\end{pmatrix} e^{-i \vect G_2\cdot \vect d_0} 
\\
&+
\begin{pmatrix} 
w_2 & w_{3} e^{i2\pi/3} \\
w_{3} e^{-i2\pi/3} & w
\end{pmatrix} e^{-i \vect G_3\cdot \vect d_0}.
\end{aligned}
\label{Td0}
\end{equation}

As mentioned in Ref.[\onlinecite{Wu_2019}] quantities roughly have orders $w_{2,3}\sim 10meV\ll\Delta_g$ and for values of $k_i$ close to the $\vect K$ point $a_0 tk_i\ll \Delta_g$. Therefore $\Delta_g$ is the dominant energy scale and the low-energy manifold is spanned by the $1/\sqrt{2}(d_{x^2-y^2}+id_{xy})$ orbitals of each layer. We can make use of any choice of downfolding procedure such as the L\"owdin partitioning method  \cite{doi:10.1002/qua.560210105}, Schrieffer-Wolf transformation  \cite{slagle2017fracton,Wurtz_2020,Bravyi_2011} or Brillouin-Wigner degenerate perturbation theory \cite{hubavc2010brillouin}. In any case one finds the Hamiltonian with the form of Eq. \eqref{Hd0}  to second order in $t$ and to first order in $w$.

From here we can directly answer the two original questions. First, light in free space described by $\vect A=(A_x,A_y,0)$ couples via the momenta $k_i\to k_i-A_i$ and  waveguide-type light $\vect A=(0,0,A_z)$ couples via $w\to e^{-iAa_L}$. This now is clear because of the connections to the underlying tight binding model. Second, one should consider light of frequencies $\Omega\ll \Delta_g$ to avoid couplings to higher energy states.

\subsection{Better single layer dispersion}
\label{better_single_layer_dispersion}

As we saw in the previous appendix the bands of a single layer TMD enter directly into the Hamiltonian via a downfolding procedure applied to bands that are valid near the $\vect K$ point. Now one may ask the question if we can do better than that in the sense that the bands are reproduced more accurately. And one way of doing so is to start from the tight binding Hamiltonian with third nearest neighbor hopping that is provided in \cite{liu2013three}.

\begin{equation}
\begin{aligned}
&H^{\text{TNN}}(\vect k)=\begin{pmatrix}V_{0} & V_{1} & V_{2}\\
V_{1}^{*} & V_{11} & V_{12}\\
V_{2}^{*} & V_{12}^{*} & V_{22}
\end{pmatrix}
\end{aligned}.
\end{equation}
This Hamiltonian is written in the basis of d-orbitals $\{d_{z^2},d_{xy},d_{x^2-y^2}\}$.
The different contributions to the Hamiltonian are given as
\begin{equation}
\begin{aligned}
V_{0}&=\varepsilon_{1}+2t_{0}(2\cos X_k\cos Y_k+\cos2X_k)\\
&+2r_{0}(2\cos3X_k\cos Y_k+\cos2Y_k)\\
&+2u_{0}(2\cos2X_k\cos2Y_k+\cos4X_k)\\
\end{aligned}
\end{equation}

\begin{equation}
\begin{aligned}
V_{1}&=-2\sqrt{3}t_{2}\sin X_k\sin Y_k\\
&+2(r_{1}+r_{2})\sin3X_k\sin Y_k\\
&-2\sqrt{3}u_{2}\sin2X_k\sin2Y_k\\
&+i\Big[2t_{1}\sin X_k(2\cos X_k+\cos Y_k)\\
&+2(r_{1}-r_{2})\sin3X_k\cos Y_k\\
&+2u_{1}\sin2X_k(2\cos2X_k+\cos2Y_k)\Big]
\end{aligned}
\end{equation}

\begin{equation}
\begin{aligned}
V_{2}&=2t_{2}(\cos2X_k-\cos X_k\cos Y_k)\\
&-\frac{2}{\sqrt{3}}(r_{1}+r_{2})(\cos3X_k\cos Y_k-\cos2Y_k)\\
&+2u_{2}(\cos4X_k-\cos2X_k\cos2Y_k)\\
&+i\Big[2\sqrt{3}t_{1}\cos X_k\sin Y_k\\
&+\frac{2}{\sqrt{3}}\sin Y_k(r_{1}-r_{2})(\cos3X_k+2\cos Y_k)\\
&+2\sqrt{3}u_{1}\cos2X_k\sin2Y_k\Big]
\end{aligned}
\end{equation}

\begin{equation}
\begin{aligned}
V_{11}&=\varepsilon_{2}+(t_{11}+3t_{22})\cos X_k\cos Y_k+2t_{11}\cos2X_k\\
&+4r_{11}\cos3X_k\cos Y_k+2(r_{11}+\sqrt{3}r_{12})\cos2Y_k\\
&+(u_{11}+3u_{22})\cos2X_k\cos2Y_k+2u_{11}\cos4X_k,
\end{aligned}
\end{equation}
\begin{equation}
\begin{aligned}
V_{12}&=\sqrt{3}(t_{22}-t_{11})\sin X_k\sin Y_k+4r_{12}\sin3X_k\sin Y_k\\
&+\sqrt{3}(u_{22}-u_{11})\sin2X_k\sin2Y_k\\
&+i\Big[4t_{12}\sin X_k(\cos X_k-\cos Y_k)\\
&+4u_{12}\sin2X_k(\cos2X_k-\cos2Y_k)\Big]
\end{aligned}
\end{equation}
 with the shorthand
\begin{equation}
	X_k=\frac{1}{2}k_xa;\quad Y_k=\frac{\sqrt{3}}{2}k_ya
\end{equation}

and couplings in units of eV are given as
\begin{equation}
	\begin{aligned}
	&\varepsilon_1=0.588;\quad\varepsilon_2=1.303;\quad t_0=-0.226\\
	&t_1=-0.234;\quad t_2=0.036;\quad t_{11}=0.400\\
	&t_{12}=0.098;\quad t_{22}=0.017;\quad r_0=0.003\\
	&r_1=-0.025;\quad r_2=-0.169;\quad r_{11}=0.082\\
	&r_{12}=0.051;\quad u_0=0.057;\quad u_1=0.103\\
	&u_2=0.187;\quad u_{11}=-0.045;\quad u_{12}=-0.141\\
	&u_{22}=0.087
	\end{aligned}.
\end{equation}
The lowest energy band of this Hamiltonian is what the band that the moir\'e bands described in \eqref{Hd0} are based on. 

The most obvious thing to do would be to blindly diagonalize the Hamiltonian and keep the expression for the lowest energy band. This expression, however, is cumbersome because it involves the solution of a cubic equation. The nested square-roots that appear in the solution would make further progress in computations difficult and is not very economical. The way we can do better is to recognize that the lattice is hexagonal and that therefore the Brillouin zone and bands have 6 fold rotational symmetry under $\vect k\to R(n\pi/3)\cdot \vect k$ and mirror symmetries $(k_x,k_y)\to (-k_x,k_y)$ and $(k_x,k_y)\to (k_x,-k_y)$. We may now start with a Fourier series
\begin{equation}
\begin{aligned}
E(\vect k)=&\sum_{n,m}(c_{nm}\cos[(n_1v_1+n_2v_2)\cdot\vect k]\\
&+s_{nm}\sin[(n_1v_1+n_2v_2)\cdot\vect k])
\end{aligned}
\end{equation}
where $v_1=(a,0)$ and $v_2=a/2(1,\sqrt{3})$ are lattice vectors. After imposing the symmetries and truncating such that $n,m=-5,\dots 5$ we find the ansatz
\begin{equation}
	\begin{aligned}
	E(\vect k)=&C_0+\sum_{n=1}^5 C_n[2 \cos (n X_k) \cos (n Y_k)+\cos (2n X_k)]\\
	&+C_6[2 \cos (3 X_k) \cos (Y_k)+\cos (2 Y_k)]\\
	&+C_7[2 \cos (6 X_k) \cos (2 Y_k)+\cos (4 Y_k)]\\
	&+2 C_8 \left[\cos (8 X_k) \cos (2 Y_k)+\cos (7 X_k) \cos (3 Y_k)\right.\\
	&+\left.\cos (X_k) \cos (5 Y_k)\right]\\
	&+2 C_9 \left[\cos (9 X_k) \cos (Y_k)+\cos (6 X_k) \cos (4 Y_k)\right.\\
	&+\left.\cos (3 X_k) \cos (5 Y_k)\right]\\
	&+2 C_{10}[\cos (7 X_k) \cos (Y_k)+\cos (5 X_k) \cos (3 Y_k)\\
	&+\cos (2 X_k) \cos (4 Y_k)]\\
	&+2 C_{11}[\cos (5 X_k) \cos (Y_k)+\cos (4 X_k) \cos (2 Y_k)\\
	&+\cos (X_k) \cos (3 Y_k)]
	\end{aligned}.
\end{equation}

Now one can fit this ansatz to the lowest energy band of $H^{\text{TNN}}$. However, one still has to be careful to include a few constraints to ensure that the low-energy  part of the spectrum is captured as accurately as possible. That is we require that
\begin{equation}
	\left.\frac{d^2E(\vect k)}{dk_idk_j}\right|_{\vect k=\vect K}=-\frac{1}{m^*}
\end{equation}
and that the value at the $\vect K$ points correctly reproduces the result from $H^{TNN}$.

\end{document}